\documentclass[aps,prb,showpacs,preprintnumbers,11pt]{revtex4-1}
\usepackage{mathrsfs}
\usepackage{amsmath}
\usepackage{amsfonts}
\usepackage{amssymb}
\usepackage{amsthm}
\usepackage{graphicx,natbib}
\usepackage{color}
\usepackage{hyperref}
\usepackage[caption=false]{subfig}

\begin{document}
\title{Coherent elastic excitation of spin waves}
\author{Akashdeep Kamra$^{1}$}
\author{Hedyeh Keshtgar$^{2}$}
\author{Peng Yan$^{1}$}
\author{Gerrit E. W. Bauer$^{3,1}$}
\affiliation{$^{1}$Kavli Institute of NanoScience, Delft University of Technology, Lorentzweg 1, 2628 CJ Delft, The Netherlands}
\affiliation{$^{2}$Institute for Advanced Studies in Basic Science, 45195 Zanjan, Iran}
\affiliation{$^{3}$Institute for Materials Research and WPI-AIMR, Tohoku University, Sendai 980-8577, Japan}

\begin{abstract}
We model the injection of elastic waves into a ferromagnetic film (F) by a non-magnetic transducer (N). We compare the configurations in which the magnetization is normal and parallel to the wave propagation. The lack of axial symmetry in the former results in the emergence of evanescent interface states. We compute the energy-flux transmission across the N$|$F interface and sound-induced magnetization dynamics in the ferromagnet. We predict efficient acoustically induced pumping of spin current into a metal contact attached to F. 
\end{abstract}

\maketitle



\section{Introduction}

The macroscopic magnetic moment of a ferromagnet results from a
symmetry-broken ground state in which the constituent spins align by the
exchange interaction.~\cite{Chikazumi1997} The underlying crystal lattice
breaks the rotational invariance of the magnetic order. Owing to spin-orbit
interaction and dipolar fields, the spins experience elastic deformations in
the form of a magneto-elastic coupling (MEC). Vice versa, the lattice is
affected by the magnetization in the form of, e.g. magnetostriction. The MEC
appears to be the dominant cause for Gilbert damping~\cite{Gilbert2004} of the
magnetization dynamics of insulators and plays the key role in equilibration
of the magnetic system with its surroundings.~\cite{Akhiezer1968} It also
offers elastic control of magnetization dynamics.

While the coupled elastic and magnetic dynamics was first investigated half a
century ago,~\cite{Akhiezer1968,Kittel1958} interest in this area has been
rekindled by improved material growth and fabrication methods. Uchida
\textit{et al.}~\cite{Uchida2011} induced spin pumping by longitudinal
acoustic waves injected into a ferromagnetic insulator, suggesting MEC to be a
possible mechanism behind the transverse spin Seebeck
effect.~\cite{Uchida2010} Weiler \textit{et al.} excited ferromagnetic
resonance (FMR) in a cobalt film by pulsed surface acoustic
waves.~\cite{Weiler2012} Static strains induce effective magnetic fields that
can be used to manipulate the magnetization.~\cite{Geprags2010} Full magnetization reversal
of a magnetic film on a cantilever by magneto-mechanical coupling has been
predicted.~\cite{Kovalev2005}

While several authors~\cite{Akhiezer1968,Kittel1958} investigated
magneto-elastic waves (MEWs) in magnetic bulk crystals, boundary conditions
and finite size effects, that are essential to understand ultrathin films and
nanostructures, have seldom been addressed.~\cite{Lecraw1965} We previously
proposed~\cite{Kamra2013} a scattering theory for MEW propagation analogous to
the Landauer-B\"{u}ttiker formalism for electronic transport in mesoscopic
systems.~\cite{Datta2005}

Here we study the excitation and propagation of MEWs in a ferromagnet by a
non-magnetic transducer that injects elastic waves into the ferromagnet. MEWs
are generated at the interface by MEC induced hybridization between the spin
and elastic waves. The mixing is resonantly enhanced around the (anti)crossing
of the spin and lattice wave dispersion relations at which fully mixed
magnon-polarons (MPs) are generated. Far from this region, the MEWs can be
considered dominantly magnonic (spin) or phononic (elastic). Because of their
mixed character, MPs can be excited by exposing the ferromagnet to sound waves.

The equations of motion for MEWs propagating in arbitrary directions are
derived in Section \ref{MEWinF}. Two special cases of interest are waves
traveling perpendicular to (Config. 1) and along (Config. 2) the equilibrium
magnetization, since they can be solved analytically and offer direct physical
insights. Here we focus on Config. 1 and compare results with Config. 2 where
appropriate.~\cite{Kamra2013} Physically, Config. 1 differs from Config. 2 by
the broken axial symmetry that causes a mixing of the right and left
precessing spin waves. We formulate the basis for a scattering matrix theory
in Section \ref{eflux} and derive magneto-elastic boundary conditions (BCs) in
Section \ref{bcs}. The energy transport across a non-magnet$|$ferromagnet
interface and the resulting excitation of MEWs are given in Section \ref{1int}.
Considering thin film ferromagnets, we investigate finite size effects such as
standing wave excitations in Section \ref{2int}. We conclude with a discussion
in Section \ref{conc}.


\section{Theoretical Method}

\subsection{Magneto-elastic waves in ferromagnets}

\label{MEWinF}

In this section, we recapitulate the continuum theory of low energy
excitations in a ferromagnet including the magneto-elastic coupling. We
closely follow Kittel~\cite{Kittel1958} to obtain the coupled equations of
motion for magnetization ($\mathbf{M}$) and displacement ($\mathbf{R}$) fields. An
applied magnetic field and easy-axis anisotropy, and thus the equilibrium
magnetization direction, is chosen along the $\hat{\mathbf{z}}$ direction (see
Fig. \ref{config}).

\subsubsection{Energy density in a ferromagnet}

The free energy density $\mathcal{H}$ has contributions from the Zeeman
interaction, magnetic anisotropy, exchange interaction, MEC, and elastic
energy:
\begin{equation} 
\mathcal{H}=\mathcal{H}_{\mathrm{Z}}+\mathcal{H}_{\mathrm{an}}+\mathcal{H}%
_{\mathrm{ex}}+\mathcal{H}_{\mathrm{MEC}}+\mathcal{H}_{\mathrm{el}}. \label{tot}
\end{equation}
For small deviations from equilibrium ($M_{x,y}\ll M_{z}\approx M_{s}$, the
saturation magnetization), Zeeman plus anisotropy energy densities
read:~\cite{Kittel1958}
\begin{equation}
\mathcal{H}_{\mathrm{Z}}+\mathcal{H}_{\mathrm{an}}=\frac{\omega_{0}}{2\gamma
M_{s}}\left(  M_{x}^{2}+M_{y}^{2}\right)  ,
\end{equation}
where $\omega_{0}=\gamma\mu_{0}H$ is the ferromagnetic resonance
frequency, $H$ is the magnitude of the external plus the anisotropy fields along $\hat{\mathbf{z}}$, $\mu_0$ is the vacuum permeability, and $\gamma(>0)$ is the gyromagnetic ratio. The exchange energy
density can be expressed as:~\cite{Kittel1949}
\begin{equation}
\mathcal{H}_{\mathrm{ex}}=\frac{A}{M_{s}^{2}}\left[  \left(  \mathbf{\nabla}M_{x}%
\right)  ^{2}+\left(  \mathbf{\nabla}M_{y}\right)  ^{2}\right]  ,
\end{equation}
in terms of the exchange constant $A$. The elastic energy density for an
isotropic solid reads:
\begin{equation}
\mathcal{H}_{\mathrm{el}}=\frac{1}{2}\rho_{F}\left(  \dot{\mathbf{R}}\cdot
\dot{\mathbf{R}}\right)  ^{2}+\frac{\lambda_{F}}{2}\left(  \sum_{i}S_{ii}\right)
^{2}+\mu_{F}\sum_{ij}S_{ij}^{2},
\end{equation}
in terms of the density $\rho_{F}$, the Lame's constants $\lambda_{F}$ and
$\mu_{F}$, and the components of the strain tensor~\cite{Lai2009} $S_{ij}=1/2(\partial R_{i}/\partial x_{j}+\partial R_{j}/\partial x_{i})$.

For cubic symmetry the MEC energy density is parametrized by the MEC constants
$b_{1,2}$ as
\begin{align}
\mathcal{H}_{\mathrm{MEC}}= &  \frac{b_{1}}{M_{s}^{2}}\sum_{i}M_{i}^{2}%
S_{ii}+\frac{b_{2}}{M_{s}^{2}}\sum_{i\neq j}M_{i}M_{j}S_{ij}+\frac{r_{0}%
}{3M_{s}^{2}}\frac{\partial A}{\partial r}\left[  \left(  \mathbf{\nabla}M_{x}%
\right)  ^{2}+\left(  \mathbf{\nabla}M_{y}\right)  ^{2}\right]  \left(  \sum
_{i}S_{ii}\right)  ,\label{mecall}\\
\approx &  \frac{2b_{2}}{M_{s}}\left(  M_{x}S_{xz}+M_{y}S_{yz}\right)
,\label{hmec}%
\end{align}
where $r$ is the distance between nearest neighbor spins with equilibrium value $r_{0}$,
and only terms linear in $M_{x,y}$ have been retained in the second step. The effects of the non-linear terms have been considered elsewhere.~\cite{Keshtgar2014} The (disregarded) last term in Eq. (\ref{mecall})
represents the MEC~\cite{Du2005} mediated by the dependence of the exchange
integral on $r$. Considering the linear terms only, we may interpret the MEC as an effective Zeeman field with
its $x$ and $y$ components proportional to $S_{xz}$ and $S_{yz}$, respectively.

\subsubsection{Equations of motion}

\begin{figure}[tb]
\begin{center}
\subfloat[Config. 1]{\includegraphics[width=70mm]{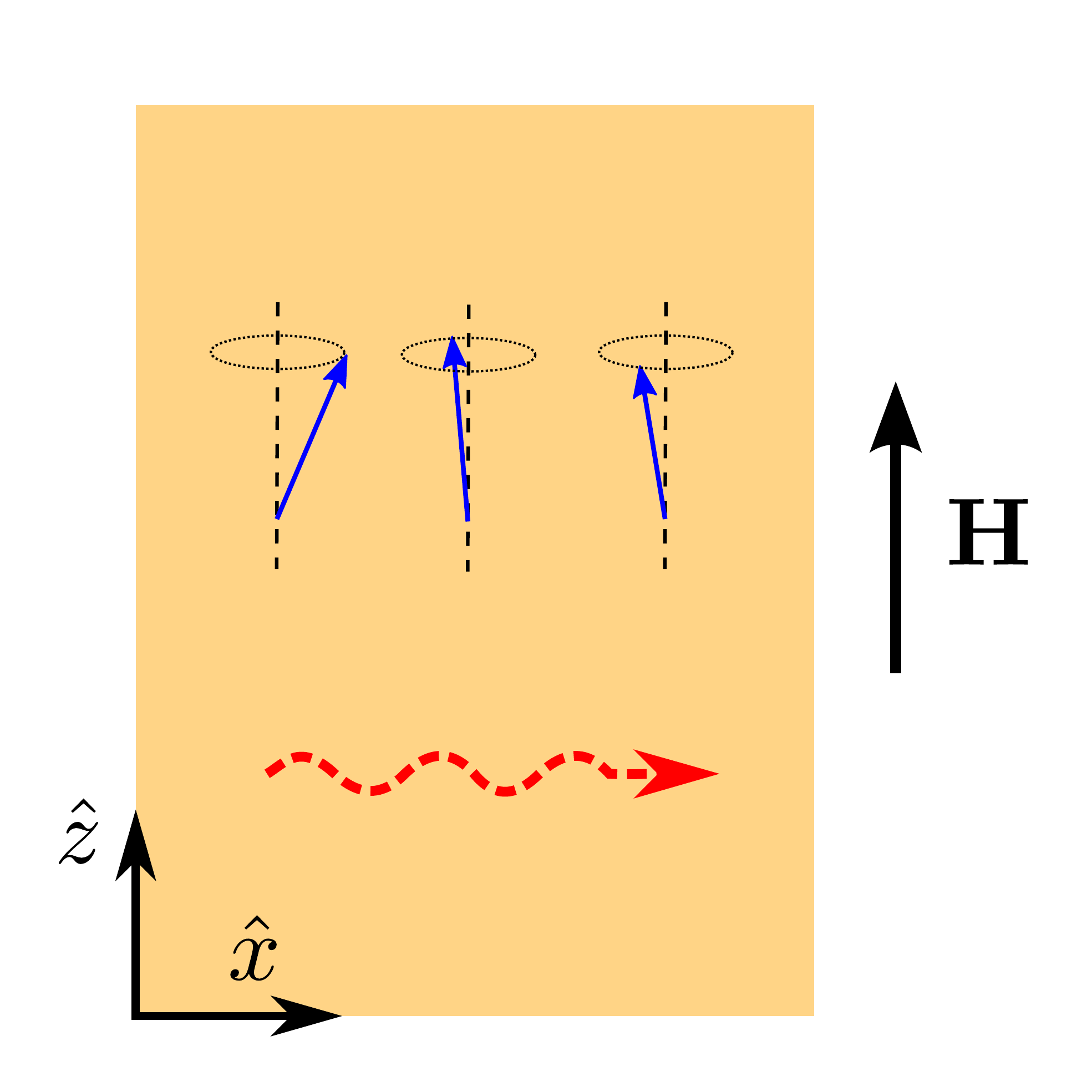}}
\subfloat[Config. 2]{\includegraphics[width=70mm]{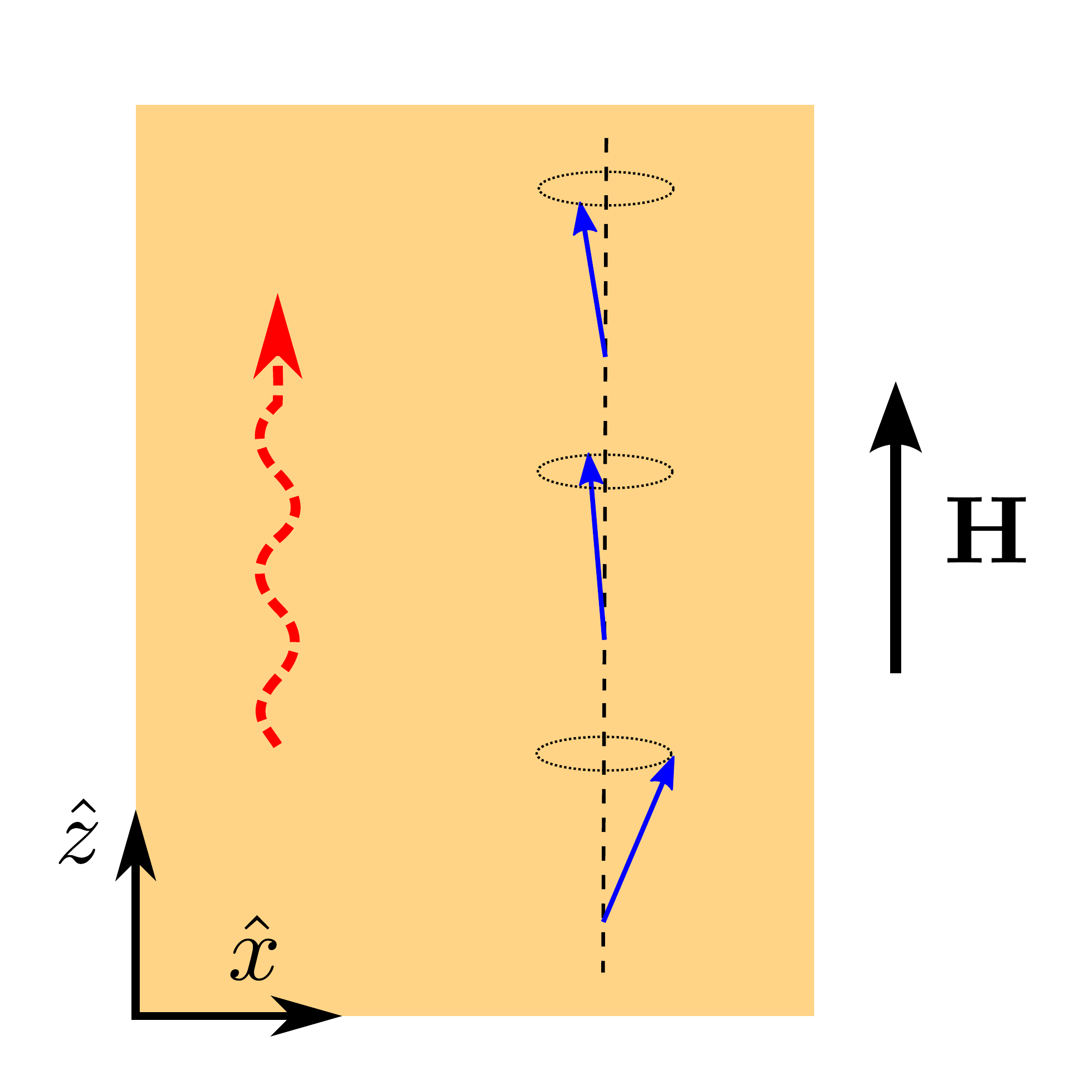}}
\end{center}
\caption[Two canonical configurations for magneto-elastic wave propagation in
ferromagnets]{Two canonical configurations for magneto-elastic wave
propagation in ferromagnets. The magnetization is saturated along the $z$ axis
by a magnetic field $\mathbf{H}$. The blue arrows depict an instantaneous magnetization texture with spin wave excitations. The red wavy arrow represents wave propagation along (a)
$\hat{\mathbf{x}}$ and (b) $\hat{\mathbf{z}}$.}%
\label{config}%
\end{figure}

The Hamilton equations of motion for the energy density defined above
read:~\cite{Kittel1958,Comstock1963}
\begin{align}
\dot{M}_{x}= &  \omega_{0}M_{y}-D\nabla^{2}M_{y}+b_{2}\gamma\left(
\frac{\partial R_{y}}{\partial z}+\frac{\partial R_{z}}{\partial y}\right)
,\label{eommx}\\
\dot{M}_{y}= &  -\omega_{0}M_{x}+D\nabla^{2}M_{x}-b_{2}\gamma\left(
\frac{\partial R_{x}}{\partial z}+\frac{\partial R_{z}}{\partial x}\right)
,\label{eommy}\\
\rho_{F}\ddot{R_{x}}= &  \mu_{F}\nabla^{2}R_{x}+(\lambda_{F}+\mu_{F}%
)\frac{\partial}{\partial x}\mathbf{\nabla}\cdot\mathbf{R}+\frac{b_{2}}{M_{s}}%
\frac{\partial M_{x}}{\partial z},\label{eomrx}\\
\rho_{F}\ddot{R_{y}}= &  \mu_{F}\nabla^{2}R_{y}+(\lambda_{F}+\mu_{F}%
)\frac{\partial}{\partial y}\mathbf{\nabla}\cdot\mathbf{R}+\frac{b_{2}}{M_{s}}%
\frac{\partial M_{y}}{\partial z},\label{eomry}\\
\rho_{F}\ddot{R_{z}}= &  \mu_{F}\nabla^{2}R_{z}+(\lambda_{F}+\mu_{F}%
)\frac{\partial}{\partial z}\mathbf{\nabla}\cdot\mathbf{R}+\frac{b_{2}}{M_{s}%
}\left(  \frac{\partial M_{x}}{\partial x}+\frac{\partial M_{y}}{\partial
y}\right)  ,\label{eomrz}%
\end{align}
where $D=2A\gamma/M_{s}$ is the spin wave stiffness. We disregard dissipation
since we are primarily interested in magnetic insulators such as yttrium iron
garnet (YIG) with very weak Gilbert and mechanical damping. The equations
above demonstrate coupling between all 5 field variables that renders an
analytic solution intractable. In the following we therefore focus on two
configurations corresponding to wave propagation orthogonal to and along the
equilibrium magnetization direction ($z$) as shown in Fig. \ref{config}.

\textit{Config. 1:} For wave propagation along the $x$ direction the partial
derivatives with respect to $y$ and $z$ in Eqs. (\ref{eommx}) - (\ref{eomrz})
vanish and only the transverse displacement $R_{z}$ couples to the
magnetization dynamics. The MEC [Eq. (\ref{hmec})] reduces to:
\begin{equation}
\mathcal{H}_{\mathrm{MEC}}=\frac{b_{2}}{M_{s}}M_{x}\frac{\partial R_{z}%
}{\partial x},
\end{equation}
which is not invariant under rotation about the $z$ direction. With constant
coefficients the equations of motion are solved by plane waves $B(x,t)=\Re
\left[  b(k,\omega)e^{i(kx-\omega t)}\right]  $ and can be written as a matrix
equation $\mathcal{A}\chi=0$:
\begin{equation}%
\begin{pmatrix}
i\omega & \omega_{m} & 0\\
-\omega_{m} & i\omega & -ib_{2}\gamma k\\
ib_{2}k/(\rho_{F}M_{s}) & 0 & \omega^{2}-\omega_{p}^{2}%
\end{pmatrix}%
\begin{pmatrix}
m_{x}\\
m_{y}\\
r_{z}%
\end{pmatrix}
=%
\begin{pmatrix}
0\\
0\\
0
\end{pmatrix}
, \label{secconf1}
\end{equation}
where $\omega_{m}=\omega_{m}(k)=\omega_{0}+Dk^{2}$ and $\omega_{p}=\omega
_{p}(k)=k\sqrt{\mu_{F}/\rho_{F}}$ are the uncoupled magnonic and phononic
dispersion relations.

\textit{Config. 2:} For waves propagating along the $z$ direction both
transverse components $R_{x}$ and $R_{y}$ couple to the magnetization
dynamics. The MEC [Eq. (\ref{hmec})] then reduces to the axially symmetric
form:
\begin{equation}
\mathcal{H}_{\mathrm{MEC}}=\frac{b_{2}}{M_{s}}\left(  M_{x}\frac{\partial
R_{x}}{\partial z}+M_{y}\frac{\partial R_{y}}{\partial z}\right)  .
\end{equation}
By the transformation $M^{\pm}=M_{x}\pm iM_{y}$ and $R^{\pm}=R_{x}\pm iR_{y}$
the $4\times4$ matrix equation is block-diagonalized into two $2\times2$
equations:~\cite{Kamra2013}
\begin{equation}
\left(
\begin{array}
[c]{cc}%
i(\omega-\sigma\omega_{m}) & \sigma\gamma b_{2}k\\
ib_{2}k/\rho_{F}M_{s} & \omega^{2}-\omega_{p}^{2}%
\end{array}
\right)  \left(
\begin{array}
[c]{c}%
m^{\sigma}\\
r^{\sigma}%
\end{array}
\right)  =0,\label{secconf2}%
\end{equation}
where $\sigma=\pm$ is a chirality index. $m^{+}$ denotes the spin waves that
precess \textquotedblleft with\textquotedblright\ the magnetic field, while
$m^{-}$ represents counter-rotating modes with frequency $\omega=-\omega
_{m}(k)$. Since $k$ is imaginary for any (positive) frequency, these waves are
always evanescent and cannot exist in the bulk of the ferromagnet. $r^{+}$
represents the right and $r^{-}$ the left circularly-polarized elastic waves.

\textit{Comparison between Configs. 1 and 2:} Since we consider waves along
symmetry directions, only the elastic shear waves couple to the magnetization
in both cases.~\footnote{For wave propagation along a non-symmetry direction,
longitudinal elastic (pressure) waves also couple to the magnetization
dynamics.} The 3 eigenmodes for Config. 1, as will be discussed in Section
\ref{bcs}, correspond to the 3 coupled variables [see Eq. (\ref{secconf1})],
and 2 eigenmodes for Config. 2 [see Eq. (\ref{secconf2})]. The right and left
precessing magnetoelastic modes are uncoupled under the axial symmetry of
Config. 2, but they become mixed when this symmetry is broken in Config. 1. The elastic
displacement $r_{z}$ then couples to the evanescent $m^{-}$ as well as the
propagating $m^{+}$ waves. This mixing is important in the \textquotedblleft
ultra-strong\textquotedblright\ coupling regime in which the rotating wave
approximation, i.e. the neglect of the $+/-$ coupling, breaks down. Typically,
the FMR frequency is much higher than the frequency equivalent of MEC
strength, and the rotating wave approximation is valid. Nevertheless, the
evanescent waves are necessary to formulate proper boundary conditions and
affect the conversion of acoustic to magnetic energy at the interfaces.

\subsubsection{Magneto-elastic eigenmodes}

\begin{figure}[tb]
\begin{center}
\subfloat[]{\includegraphics[width=80mm]{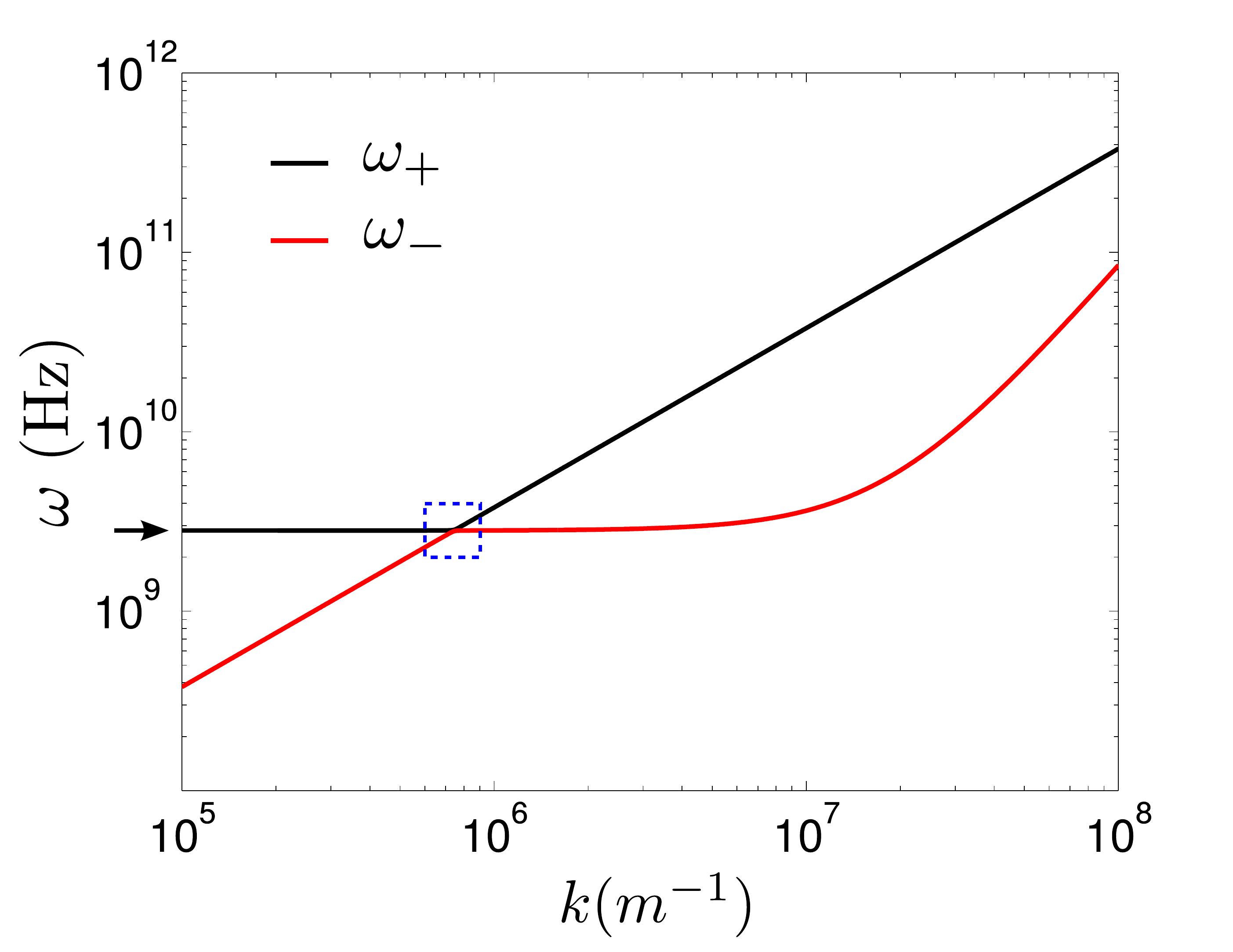}}
\subfloat[]{\includegraphics[width=76mm]{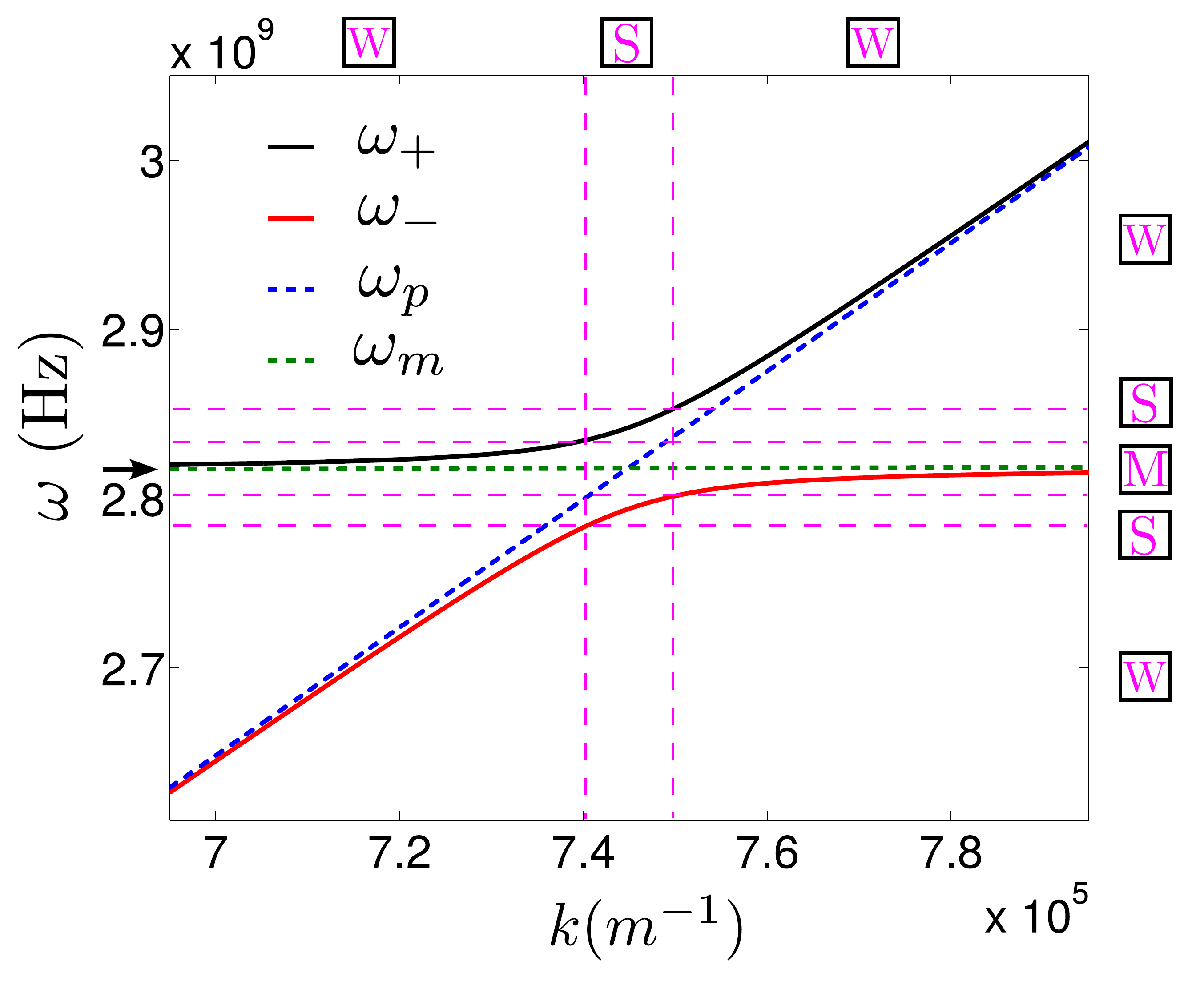}}
\end{center}
\caption[Magneto-elastic waves dispersion relation]{(a) Dispersion relation for
magneto-elastic waves (MEWs) in a ferromagnet calculated using Eq.
(\ref{drelation}) and parameters for YIG. The arrow on the ordinate indicates
the FMR frequency $\omega_{0}$. The blue dashed box is expanded in (b) to
reveal the anti-crossing. The dashed lines denote the unperturbed magnonic
($\omega_{m}$, green line) and phononic ($\omega_{p}$, blue line) dispersion
relations, while the solid lines represent the coupled MEWs. The $\omega$-$k$
space can be classified into 3 regions - (i) W region, where the MEWs can be
considered quasi-phononic or quasi-magnonic, (ii) S region, where the MEWs
have a mixed character, (iii) M region, where both excitations are
quasi-magnonic leading to a pseudo-bandgap for quasi-phononic excitations.}%
\label{disp}%
\end{figure}

Diagonalization of Eq. (\ref{secconf1}) leads to the dispersion relations of
the magneto-elastic waves (MEWs):
\begin{equation}
\omega_{\pm}=\sqrt{\frac{\omega_{m}^{2}+\omega_{p}^{2}}{2}\pm\sqrt{\left(
\frac{\omega_{m}^{2}-\omega_{p}^{2}}{2}\right)  ^{2}+\frac{b_{2}^{2}%
k^{2}\gamma\omega_{m}}{\rho_{F}M_{s}}}}. \label{drelation}
\end{equation}
The wave vectors $k_{a}$ of the eigenmodes at frequency $\omega$, where the
subscript $a$ labels the eigenmodes, are obtained by inverting the dispersion
relation [Eq. (\ref{drelation})] as will be discussed in Section \ref{bcs}.
The corresponding eigenvectors $\chi_{a}$ are:
\begin{equation}
\chi_{a}=%
\begin{pmatrix}
m_{x}\\
m_{y}\\
r_{z}%
\end{pmatrix}
=N_{a}%
\begin{pmatrix}
ib_{2}\gamma k_{a}\omega_{ma}/(\omega^{2}-\omega_{ma}^{2})\\
b_{2}\gamma k_{a}\omega/(\omega^{2}-\omega_{ma}^{2})\\
1
\end{pmatrix}
, \label{emodes}
\end{equation}
where $N_{a}$ is a dimensionless normalization factor, $\omega_{ma}%
\equiv\omega_{m}(k_{a})$, and the eigenmodes consist of elliptical
magnetization precession around $z$ coupled with the elastic shear mode along
$z$. The dispersion [Eq. (\ref{drelation})] is plotted in Fig. \ref{disp} for
parameters appropriate for YIG: $M_{s}=1.4\times10^{5}\,$A/m, $b_{2}%
=5.5\times10^{5}\,\mathrm{J}$/$\mathrm{m}^{3}$, $D=8.2\times10^{-6}%
\,\mathrm{m}^{2}$/$\mathrm{s}$, $H=8\times10^{4}~\mathrm{A/m}$, $\gamma
=2.8\times10^{10}\,\mathrm{Hz/T}$, $\rho_{F}=5170\,\mathrm{kg/m}^{3}$, and
$\mu_{F}=74\,\mathrm{GPa}$.~\cite{Eggers1963,Hansen1973,Deltronic} When the
MEC is weak, the branches in the dispersion diagram are quite close to the
uncoupled dispersion relations $\omega_{m}$ or $\omega_{p}$ [see Eq.
(\ref{drelation})] in much of phase space [W regions in Fig. \ref{disp}(b)]. {  Here the mode with frequencies close to $\omega_m$ ($\omega_p$) is dominantly magnonic (phononic).}
In the crossing regime i.e. when $4b_{2}^{2}k^{2}\gamma\omega_{m}/\rho_{F}%
M_{s}\gtrsim(\omega_{m}^{2}-\omega_{p}^{2})^{2}$, the excitations hybridize [S region in Fig. \ref{disp}(b)]. {  We refer to the quasi-particle close to $k_{0}$ at which the uncoupled dispersions cross as ``magnon-polaron (MP)''.} Since the uncoupled magnon dispersion is very flat compared to that of the phonons, we may define a narrow M region in $\omega$ space [Fig.
\ref{disp} (b)] in which the magnon character dominates both excitation modes,
while the phonon character is suppressed, {  leading to a pseudo-band gap for quasi-phononic excitations} [see also Fig \ref{sintflux} (b)].
For spin wave stiffness $D\ll\omega_{0}/k_{0}^{2}$, $k_{0}\approx\omega
_{0}\sqrt{\rho_{F}/\mu_{F}}$ and the M region covers the frequency interval
$|\omega-\omega_{0}|\lesssim\sqrt{b_{2}^{2}\omega_{0}\gamma/4\mu_{F}M_{s}%
}\approx24\,$MHz.


\subsection{Energy flux and eigenmode normalization}\label{eflux} 

Energy conservation can be expressed by the continuity
equation:~\cite{Akhiezer1968}
\begin{equation}
\frac{\partial\mathcal{H}}{\partial t}+\mathbf{\nabla}\cdot\mathbf{F} = 0,
\end{equation}
where the energy flux $\mathbf{F}=F_{x}\hat{\mathbf{x}}$:
\begin{equation}
F_{x}=-\left[  \frac{2A}{M_{s}^{2}}\left(  \frac{\partial M_{x}}{\partial
t}\frac{\partial M_{x}}{\partial x}+\frac{\partial M_{y}}{\partial t}%
\frac{\partial M_{y}}{\partial x}\right)  +\frac{\partial R_{z}}{\partial
t}\left(  \mu_{F}\frac{\partial R_{z}}{\partial x}+b_{2}\frac{M_{x}}{M_{s}%
}\right)  \right]  . \label{fluxgen}
\end{equation}
For real $k$, $B(x,t)=(b(k,\omega)e^{i(kx-\omega t)}+b^{\ast}(k,\omega
)e^{-i(kx-\omega t)})/2$, whence the time-averaged energy flux $\bar{\mathbf{F}}$
is constant:
\begin{equation}
\bar{F}_{x}=\frac{A\omega k}{M_{s}^{2}}\left(  |m_{x}|^{2}+|m_{y}|^{2}\right)
+\frac{\mu_{F}\omega k}{2}|r_{z}|^{2}+\frac{b_{2}\omega}{2M_{s}}\Im
(r_{z}^{\ast}m_{x}). \label{fluxmew}
\end{equation}
For the eigenmode $\chi_{a}$:
\begin{equation}
\bar{F}_{x}^{a}=N_{a}^{2}\left[  \frac{\mu_{F}k_{a}\omega}{2}+\frac{Ab_{2}%
^{2}\gamma^{2}}{M_{s}^{2}}\frac{k_{a}^{3}\omega(\omega^{2}+\omega_{ma}^{2}%
)}{(\omega^{2}-\omega_{ma}^{2})^{2}}+\frac{b_{2}^{2}\gamma}{2M_{s}}\frac
{k_{a}\omega\omega_{ma}}{\omega^{2}-\omega_{ma}^{2}}\right]  . \label{fluxchia}
\end{equation}
$\bar{F}$ vanishes for imaginary $k$ i.e. evanescent waves that store, but not
propagate energy. Eqs. (\ref{fluxmew})$\,$-$\,$(\ref{fluxchia}) reduce to the
flux carried by purely elastic (spin) waves in the limit $b_{2}\rightarrow0$
and $\omega\rightarrow\omega_{p}(\omega_{m})$. In transport theory it is
convenient to choose the normalization factors $N_{a}$ such that each
eigenmode carries unit energy flux, i.e. $\bar{F}_{x}^{a}=1~\mathrm{W}%
/\mathrm{m}^{2}$ [Eq. (\ref{fluxchia})]. When interested in the amplitude or
the energy density, choosing a normalization factor of $N_{a}=1$ may be
simpler. The calculated physical quantities are of course independent of the
normalization chosen. In the following we will employ flux normalized representation for the propagating waves.


\subsection{Boundary conditions and acoustic actuation of MEWs}

\label{bcs} We so far discussed MEWs in the bulk of a ferromagnet. Next,
we derive the interface connection rules for a non-magnetic transducer (N)
attached to a ferromagnet (F). The required boundary conditions (BCs) can be
obtained by integrating the equations of motion over the abrupt interface with
discontinuous constitutive parameters. This is equivalent to demanding
continuity of the energy flux [Eq. (\ref{fluxgen})] across the
interface.~\cite{Akhiezer1968} The first BC corresponds to zero spin wave
angular momentum flux at the interface or ``free'' BC for the magnetization:
\begin{equation}
\left.  \frac{\partial M_{x,y}}{\partial x}\right\vert _{F}=0. \label{bc12}
\end{equation}
{  Here, we disregard the anisotropies that could ``pin'' the magnetization at the interface.} Continuity of mass velocity (or equivalently, displacement) at the
interface implies
\begin{equation}
\left.  \frac{\partial R_{z}}{\partial t}\right\vert _{F}=\left.
\frac{\partial R_{z}}{\partial t}\right\vert _{N}. \label{bc3}
\end{equation}
The third BC is the continuity of stress at the interface:
\begin{equation}
\left.  \left(  \mu_{F}\frac{\partial R_{z}}{\partial x}+b_{2}\frac{M_{x}%
}{M_{s}}\right)  \right\vert _{F}=\left.  \mu_{N}\frac{\partial R_{z}%
}{\partial x}\right\vert _{N}. \label{bc4}
\end{equation}
These BCs should be satisfied for all frequencies. The wave numbers $k_{a}$ in
F corresponding to a given frequency $\omega~(>0)$ of the elastic wave
incident from N are obtained by inverting the MEW dispersion relation [Eq.
(\ref{drelation})]. The secular equation
\begin{align}
0= &  \frac{\mu_{F}D^{2}}{\rho_{F}}k^{6}+\left[  \frac{2\omega_{0}D\mu_{F}%
}{\rho_{F}}-\frac{b_{2}^{2}\gamma D}{\rho_{F}M_{s}}-D^{2}\omega^{2}\right]
k^{4}\nonumber \\
&  -\left[  2\omega^{2}\omega_{0}D+\frac{\mu_{F}}{\rho_{F}}(\omega^{2}%
-\omega_{0}^{2})+\frac{b_{2}^{2}\gamma\omega_{0}}{\rho_{F}M_{s}}\right]
k^{2}+\omega^{2}(\omega^{2}-\omega_{0}^{2}), \label{ksol}
\end{align}
is cubic in $k^{2}$ implying 3 (doubly degenerate) solutions. One of these
solutions ($k_{1}$) is real for all $\omega$, representing a propagating wave.
It corresponds to the $\omega_{-}$ branch of the dispersion [Fig. \ref{disp}
(a)] with limiting values $k_{1}\rightarrow\omega\sqrt{\rho_{F}/\mu_{F}}$ for
$\omega<\omega_{0}$ and $k_{1}\rightarrow\sqrt{(\omega-\omega_{0})/D}$ for
$\omega>\omega_{0}$ for $b_{2}\rightarrow0$. The second root $k_{2}$
corresponds to the upper $\omega_{+}$ branch of the dispersion with, for
$b_{2}\rightarrow0$, limiting values $k_{2}\rightarrow\sqrt{(\omega
-\omega_{0})/D}$ for $\omega\lesssim\omega_{0}$ and $k_{2}\rightarrow
\omega\sqrt{\rho_{F}/\mu_{F}}$ for $\omega>\omega_{0}$, therefore evanescent
for $\omega$ below and propagating above $\omega_{0}$. The third solution
$k_{3}\rightarrow i\sqrt{(\omega+\omega_{0})/D}$ in the limit $b_{2}%
\rightarrow0$ is always evanescent and thus does not appear in the dispersion
diagram [Fig. \ref{disp} (a)].


\section{Results}

\subsection{Acoustic energy transfer across N$|$F interfaces}

\label{1int}

\begin{figure}[tb]
\begin{center}
\includegraphics[width=85mm]{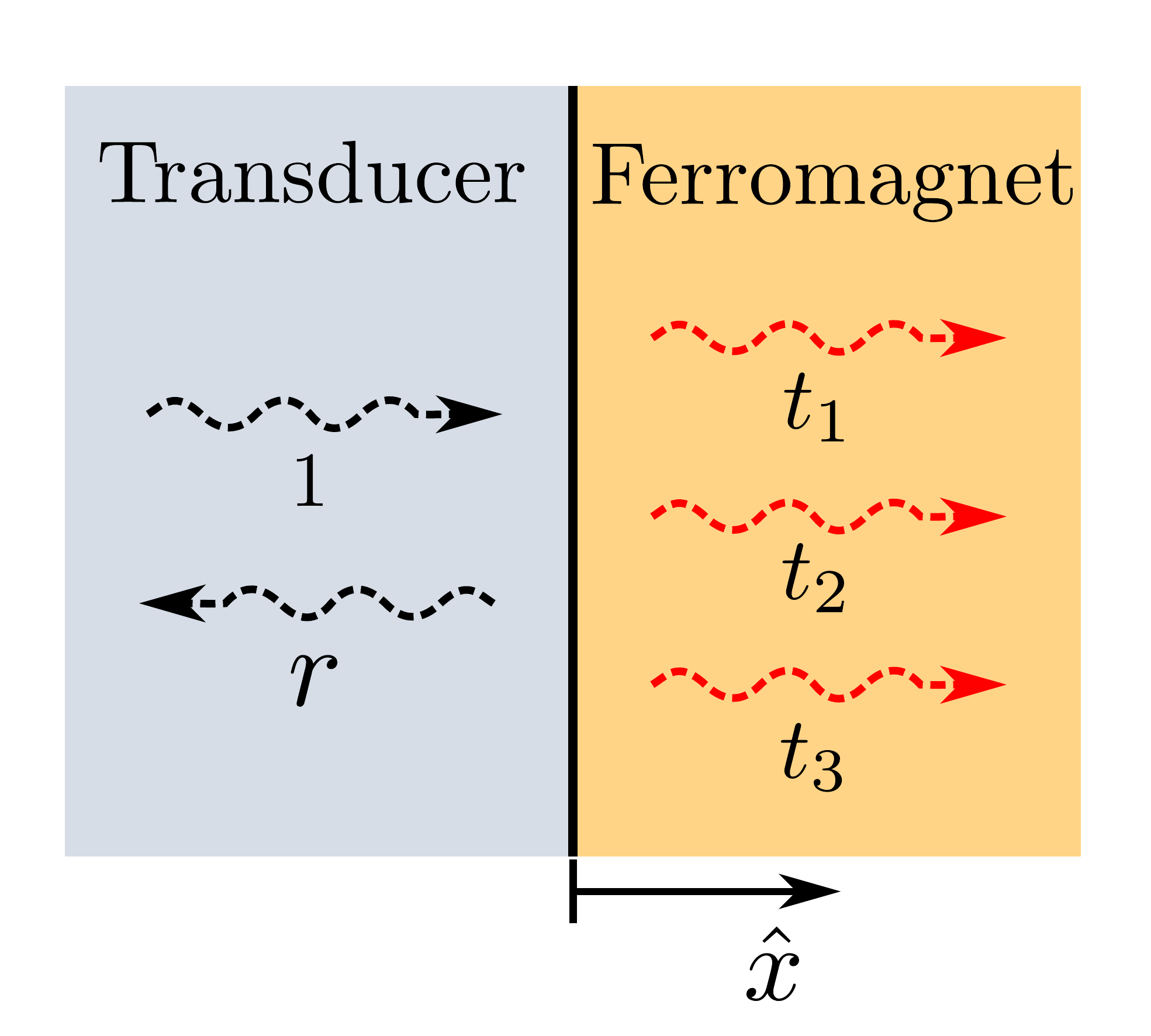}
\end{center}
\caption[Energy transport across N$|$F interface]{Scattering process at an
N$|$F interface with magnetization along $\hat{\mathbf{z}}$. Linear polarized
(along $\hat{\mathbf{z}}$) transverse acoustic waves generated in a non-magnetic
transducer (N) impinge on a ferromagnet (F) in the $x$-direction. The incident
wave is partially reflected (without mode conversion) and partially
transmitted into the F as three MEWs (shown as red wavy arrows). One of these
MEWs is always propagating, the second one is evanescent or traveling
depending on the frequency of the incident wave, while the third one is always
evanescent.}%
\label{singleint}%
\end{figure}

Here we consider a ferromagnet (F) in contact with a non-magnetic transducer
(N) that injects elastic waves propagating along $\hat{\mathbf{x}}$ (see Fig.
\ref{singleint}). Both F and N are semi-infinite (or with a perfect absorber
attached to the F side) so that only the N$|$F interface at $x=0$ matters
while there are no incoming propagating waves from F. With these boundary
conditions a flux-normalized sound wave in N (with parameters denoted by
subscript $N$) reads
\begin{eqnarray}
\psi_{N}(x\leq0) = \begin{pmatrix}
M_x, & M_y, & R_z
\end{pmatrix}
^{\intercal} & = & \sqrt{\frac{2}{\mu_{N}\omega k_{i}}}%
\begin{pmatrix}
0, & 0, & 1
\end{pmatrix}
^{\intercal}e^{i(k_{i}x-\omega t)} \nonumber \\
  & & +r(\omega)\sqrt{\frac{2}{\mu_{N}\omega
k_{i}}}%
\begin{pmatrix}
0, & 0, & 1
\end{pmatrix}
^{\intercal}e^{-i(k_{i}x+\omega t)}, \label{waveN}
\end{eqnarray}
where $k_{i}=\omega\sqrt{\rho_{N}/\mu_{N}}$ is the wavenumber of the incident
(and reflected) wave and $r(\omega)$ is the reflection coefficient calculated
below. In F we have to consider the three MEWs derived above:
\begin{equation}
\psi_{F}(x\geq0)=\sum_{l=1,2,3} \ t_{l}(\omega) \ \chi_{l}e^{i(k_{l}x-\omega t)}. \label{psifbc1}
\end{equation}
The propagating waves in Eq. (\ref{psifbc1}) are assumed to be flux-normalized
such that the reflection and transmission probabilities of the propagating
waves are simply given by $|r|^{2}$ and $|t_{j}|^{2}$ leading to:
\begin{equation}
|r|^{2}+\sum_{j}|t_{j}|^{2}=1,
\end{equation}
where the index $j$ runs over propagating modes only. The normalization factor
$N_{a}$ has been chosen to be 1 for evanescent modes.

\begin{figure}[tb]
\begin{center}
\subfloat[]{\includegraphics[width=80mm]{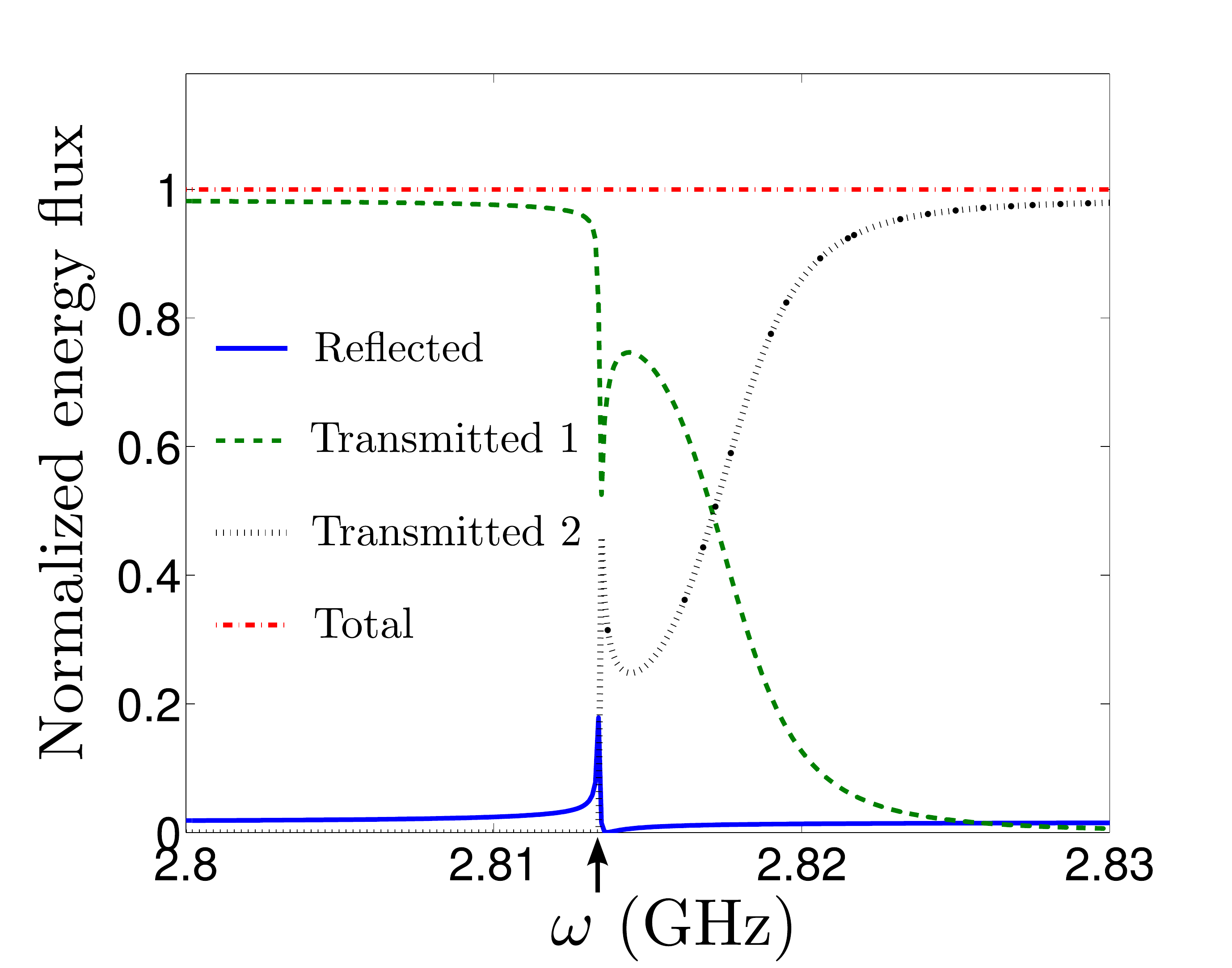}}
\subfloat[]{\includegraphics[width=78mm]{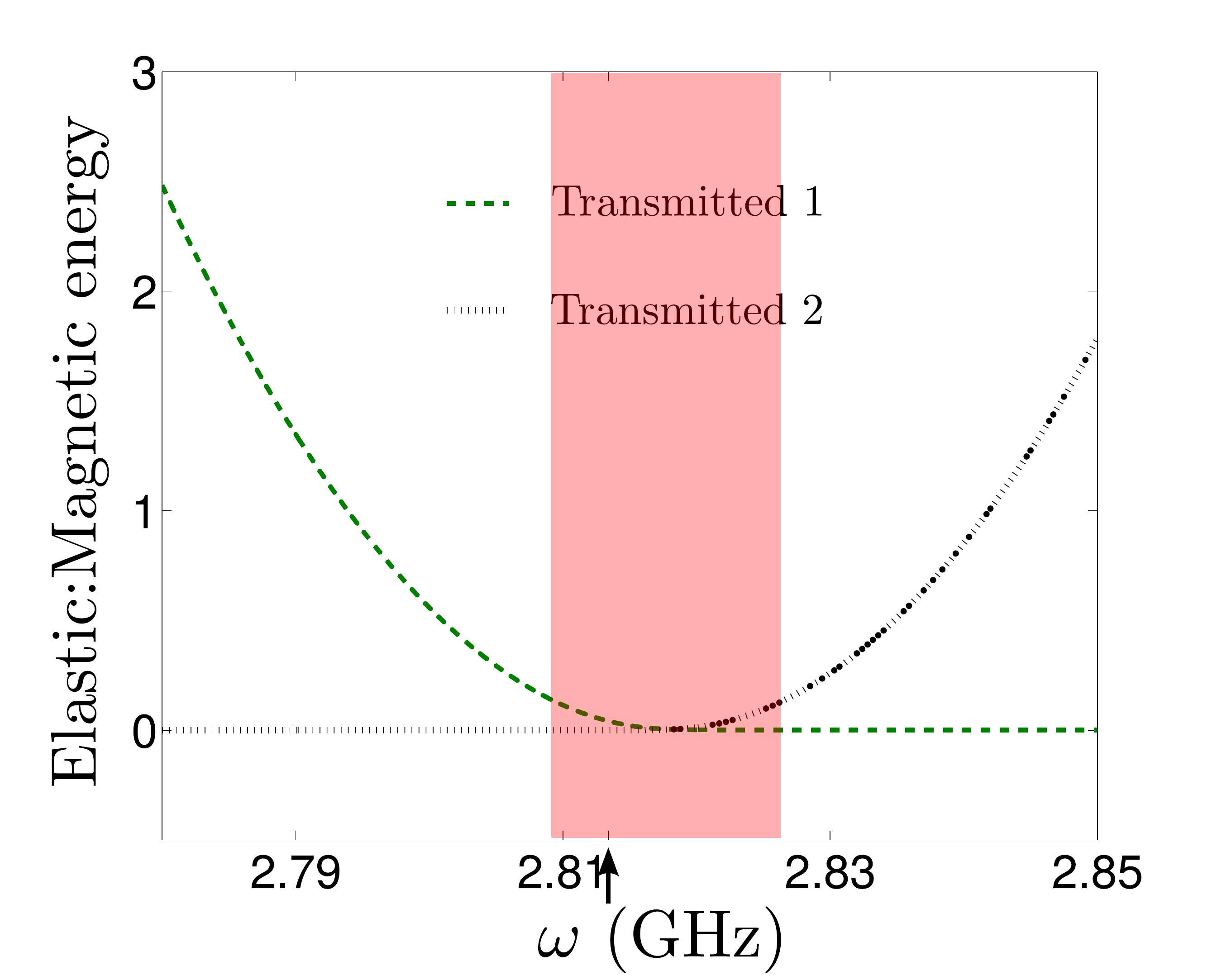}}\newline%
\subfloat[]{\includegraphics[width=80mm]{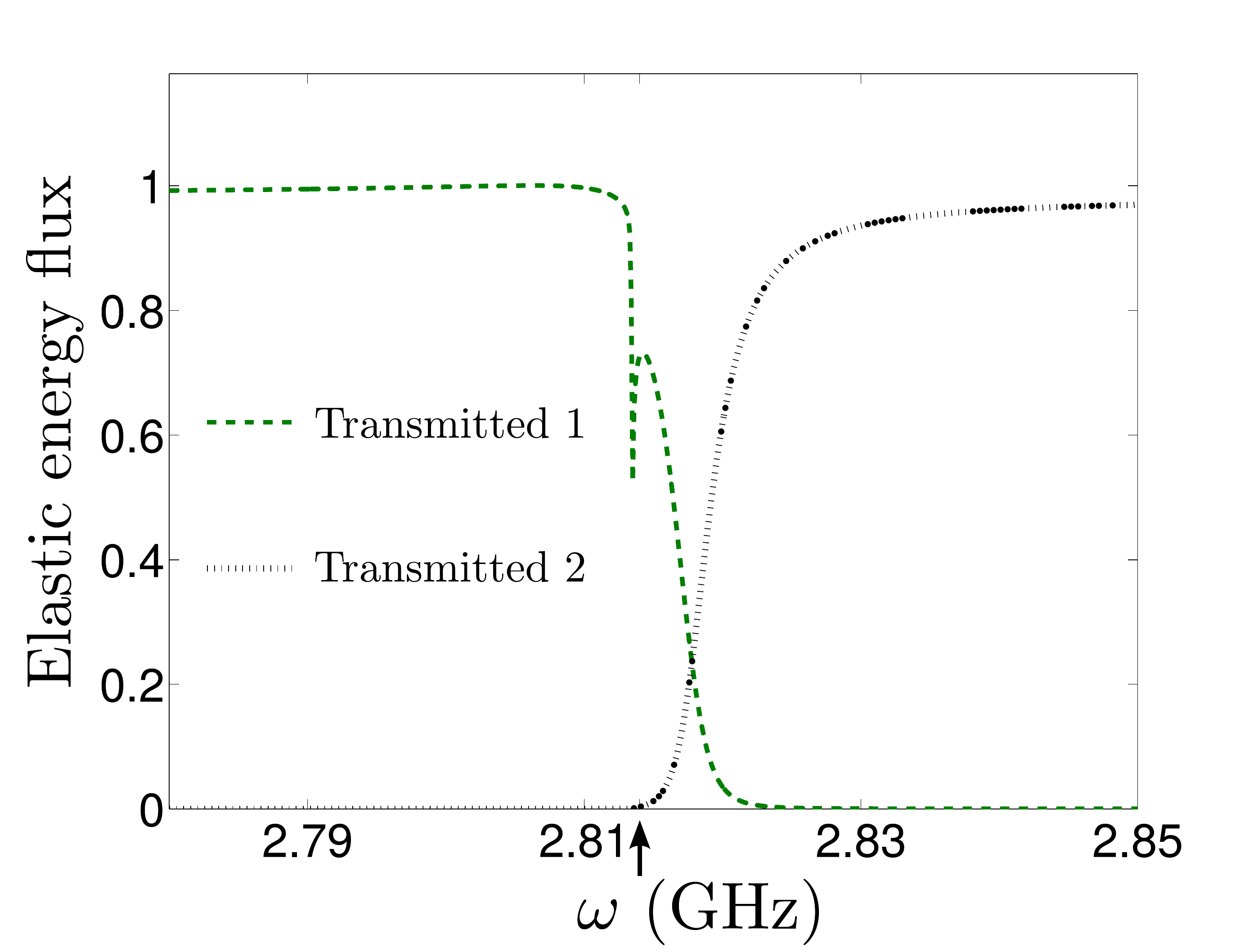}}
\end{center}
\caption[Energy flux transmission across a N$|$F interface]{(a) Normalized
energy flux carried by the reflected and transmitted waves. At $\omega_{0}$
wave 2 changes character from evanescent to propagating. (b) Ratio between the
elastic and magnetic energy densities associated with the transmitted waves.
The ratio becomes very small when the magnetic energy dominates but never
vanishes. The shaded region depicts the pseudo-bandgap for elastic waves. (c)
Lattice contribution to the energy flux for the two transmitted waves. The
arrows on the abscissas indicate the FMR frequency $\omega_{0}$.}%
\label{sintflux}%
\end{figure}

Imposing the four boundary conditions [Eqs. (\ref{bc12}) - (\ref{bc4})] yields
four equations for the four variables $r,t_{1,2,3}$:
\begin{align}
\sum_{l}t_{l}k_{l}\chi_{l}[1] &  =0,\\
\sum_{l}t_{l}k_{l}\chi_{l}[2] &  =0,\\
\sum_{l}t_{l}\chi_{l}[3] &  =\sqrt{\frac{2}{\mu_{N}\omega k_{i}}}\left(
1+r\right)  ,\\
\sum_{l}\left(  i\mu_{F}k_{l}t_{l}\chi_{l}[3]+\frac{b_{2}}{M_{s}}t_{l}\chi
_{l}[1]\right)   &  =i\mu_{N}k_{i}\sqrt{\frac{2}{\mu_{N}\omega k_{i}}}\left(
1-r\right)  ,
\end{align}
with $\chi_{l}[m]$ denoting the \textit{m}th element of the vector $\chi_{l}$.
The analytic solutions are unwieldy and not presented here. In Fig.
\ref{sintflux} (a) we plot the energy flux carried by the propagating waves
for a junction of magnetic YIG and non-magnetic gadolinium gallium garnet
(GGG) with parameters: $\rho_{N}=7085\,\mathrm{kg}$/$\mathrm{m}^{3}$, and
$\mu_{N}=90\,$\textrm{GPa}.~\cite{Graham1970} The small but finite acoustic
mismatch causes partial reflection even far from the resonance without
actuating the magnetization.

Figure \ref{sintflux} (a) is very similar to the analogous plot for the
symmetric configuration (Config. 2) considered in Ref. \onlinecite{Kamra2013} in which
circularly polarized MEWs propagate along the equilibrium magnetization
direction. {  Far from the anti-crossing, transmission is efficient into the quasi-phononic excitation. The modes gradually change their character when approaching the anti-crossing. Hence transmission into one branch increases at the cost of the other one.} The evanescent modes apparently do not affect the steady state
transmission even close to the anticrossing. We expect them to play a
significant role in the transmitted energy current only in the ultra-strong
coupling regime in which the MEC is of the order of $\omega_{0}$. However,
this does not imply that the evanescent states may be neglected. They do store
significant energy and should show up in the transients when actuation is
carried out by ultrashort pulses. Furthermore, the presence of defects would
mix the evanescent interface states with propagating ones. The MEWs are
efficiently excited in the full frequency range [see Fig. \ref{sintflux} (a)]
including MPs, which are formed at about 2.79 and 2.84 GHz [see Fig.
\ref{sintflux} (b)], although in contrast to the energy density in Fig.
\ref{sintflux} (b), the energy flux is still dominated by the lattice degree
of freedom [see Fig. \ref{sintflux} (c)].


\subsection{Excitation of spin waves in ferromagnetic films}

\label{2int}

\begin{figure}[tb]
\begin{center}
\includegraphics[width=85mm]{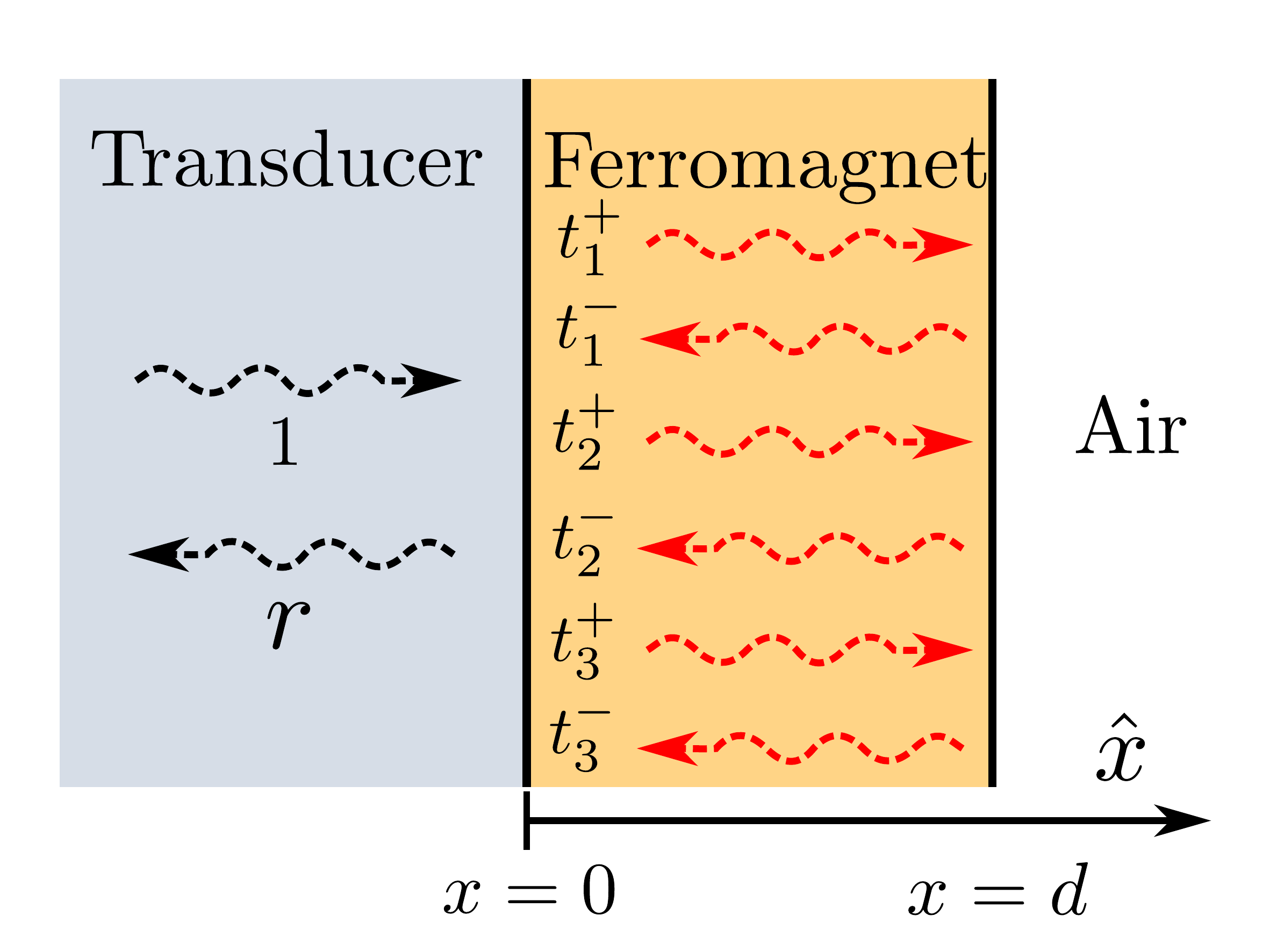}
\end{center}
\caption[Bilayer structure for excitation of spin waves in ferromagnetic
films]{Schematic of an N$|$F structure exposed to vacuum/air on the F side. Elastic waves
incident from N excite MEWs in F. The F$|$vacuum interface at $x=d$ totally
reflects all waves. Standing wave solutions in F are broadened by the energy
leakage back into N.}%
\label{doubleint}%
\end{figure}

\begin{figure}[tb]
\begin{center}
\subfloat[]{\includegraphics[height=80mm]{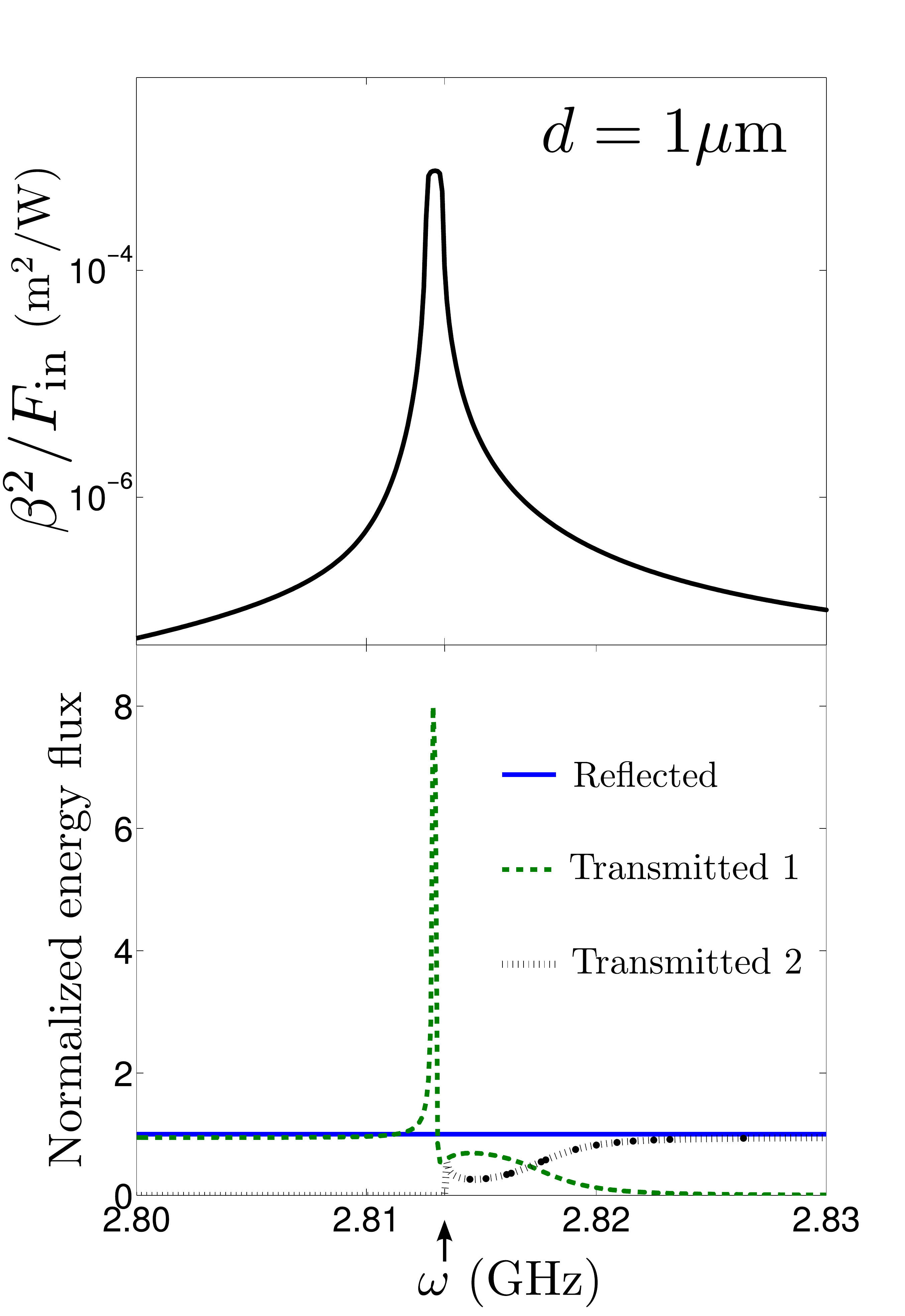}}
\subfloat[]{\includegraphics[height=80mm]{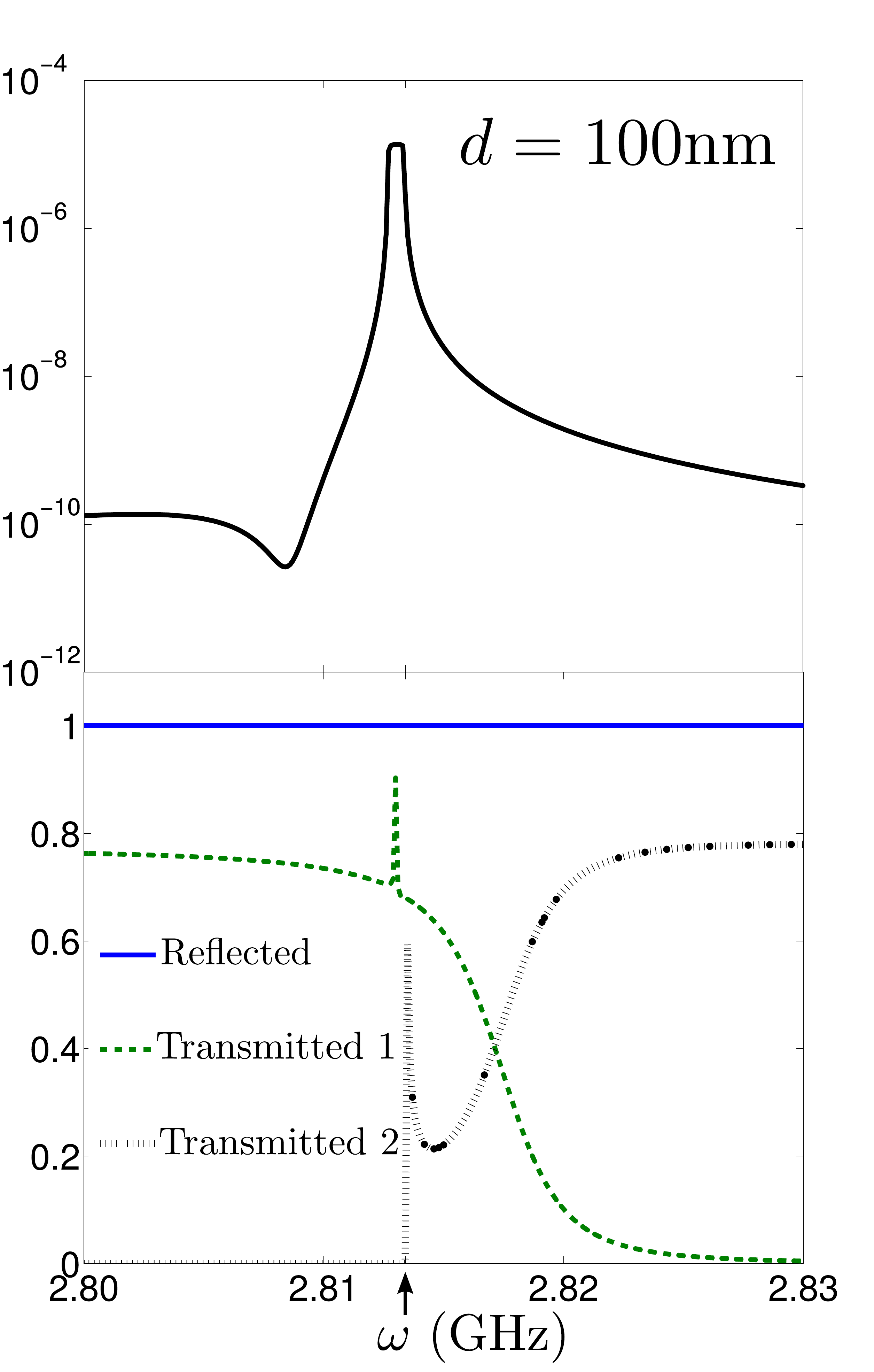}}
\subfloat[]{\includegraphics[height=80mm]{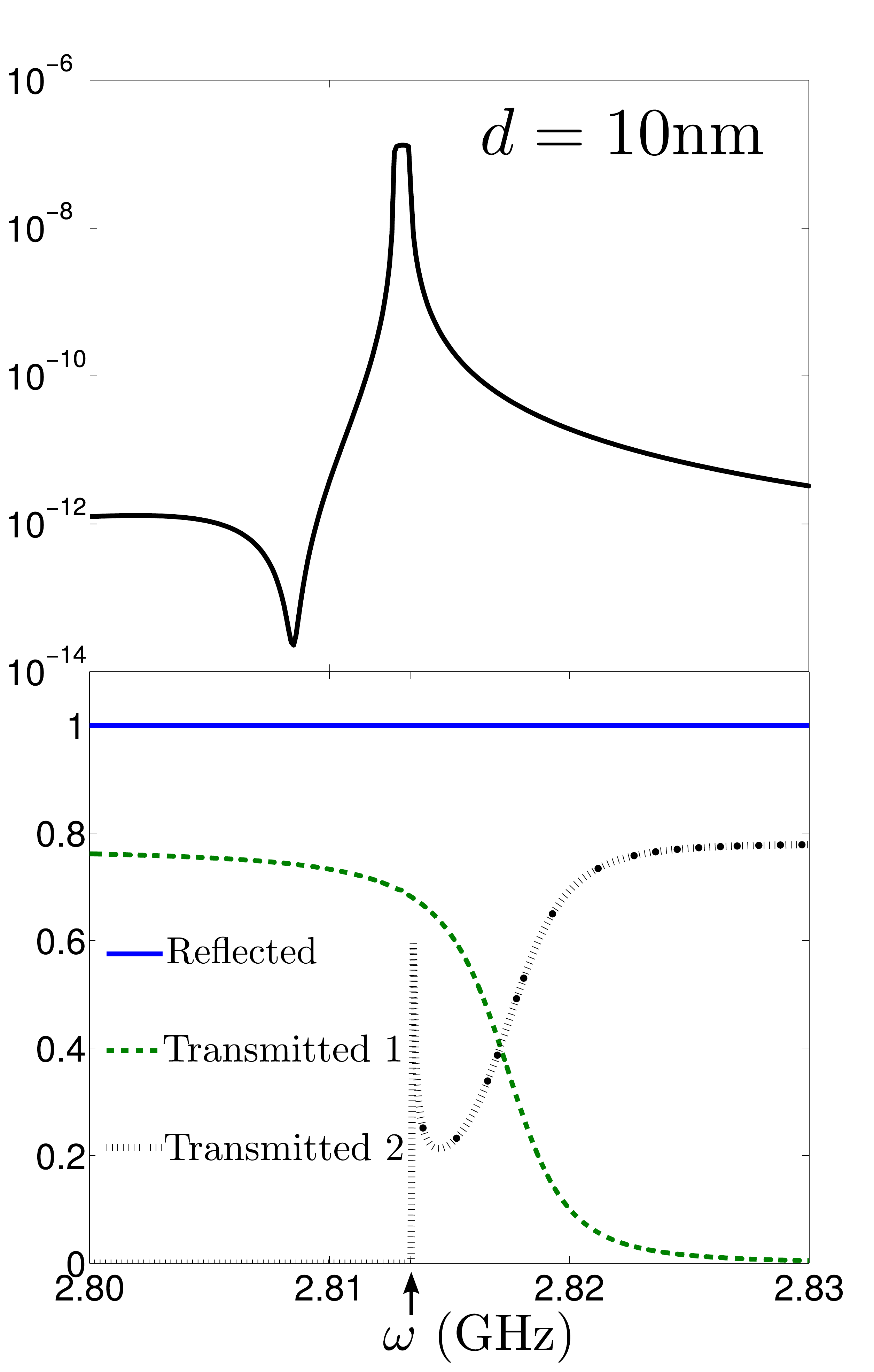}}
\end{center}
\caption[Energy fluxes and squared spin pumping angle per incident flux for different
thicknesses of the ferromagnetic layer]{Energy fluxes and squared spin pumping angle
$\beta^2$ per incident flux $F_{\mathrm{in}}$ \textit{vs.} frequency $\omega$ of
the incident elastic wave for F layer thickness (a) $d=1\,\mathrm{\mu m}$,
(b) $d=100\,$nm, and (c) $d=10\,$nm. Note the different scales on the
ordinate. The arrows on the abscissas indicate the FMR frequency $\omega_{0}$.
The fluxes shown here are carried by the forward propagating transmitted waves
in F and reflected wave in N.}%
\label{fluxspang}%
\end{figure}

We now consider finite size effects in the device depicted in Fig.
\ref{doubleint} in which F is bounded by the actuator on one side and
air/vacuum on the other. Since we disregard damping, net energy transport
through any cross section vanishes. N is still described by Eq. (\ref{waveN})
while in F:
\begin{equation}
\psi_{F}(0\leq x\leq d)=\sum_{l=1,2,3}\left(  t_{l}^{+}\chi_{l}(k_{l}%
)e^{i(k_{l}x-\omega t)}+t_{l}^{-}\chi_{l}(-k_{l})e^{-i[k_{l}(x-d)+\omega
t]}\right)  , \label{psif2}
\end{equation}
with $k_{l}>0$ for traveling and $\Im(k_{l})>0$ for evanescent waves. Since
$\chi_{l}[1,2](-k_{l})=-\chi_{l}[1,2]$ and $\chi_{l}[3](-k_{l})=\chi_{l}[3]$,
the boundary conditions [Eqs. (\ref{bc12}) to (\ref{bc4})] at $x=0$ read:
\begin{align}
\sum_{l}\left(  t_{l}^{+}k_{l}\chi_{l}[1]+t_{l}^{-}k_{l}\chi_{l}[1]e^{ik_{l}%
d}\right)   &  =0,\label{bc01}\\
\sum_{l}\left(  t_{l}^{+}k_{l}\chi_{l}[2]+t_{l}^{-}k_{l}\chi_{l}[2]e^{ik_{l}%
d}\right)   &  =0,\\
\sum_{l}\left(  t_{l}^{+}\chi_{l}[3]+t_{l}^{-}\chi_{l}[3]e^{ik_{l}d}\right)
&  =\sqrt{\frac{2}{\mu_{N}\omega k_{i}}}\left(  1+r\right)  ,\\
\sum_{l}\left[  \left(  i\mu_{F}k_{l}\chi_{l}[3]+\frac{b_{2}}{M_{s}}\chi
_{l}[1]\right)  \left(  t_{l}^{+}-t_{l}^{-}\ e^{ik_{l}d}\right)  \right]   &
=i\mu_{N}k_{i}\sqrt{\frac{2}{\mu_{N}\omega k_{i}}}\left(  1-r\right)  .
\end{align}
The total reflection corresponds to free boundary condition at the outer
interface ($x=d$):
\begin{align}
\sum_{l}k_{l}\chi_{l}[1]\left(  t_{l}^{+}e^{ik_{l}d}+t_{l}^{-}\right)   &
=0,\\
\sum_{l}k_{l}\chi_{l}[2]\left(  t_{l}^{+}e^{ik_{l}d}+t_{l}^{-}\right)   &
=0,\\
\sum_{l}\left[  \left(  i\mu_{F}k_{l}\chi_{l}[3]+\frac{b_{2}}{M_{s}}\chi
_{l}[1]\right)  \left(  t_{l}^{+}e^{ik_{l}d}-t_{l}^{-}\right)  \right]   &
=0,\label{bcd4}%
\end{align}
thereby completing the set of 7 linear equations [Eqs. (\ref{bc01}) to
(\ref{bcd4})] for 7 variables ($r,t_{l}^{\sigma}$). While no energy is
transported it is instructive to plot the normalized energy flux carried by
the reflected and the forward traveling waves, corresponding to the
coefficients $r$, $t_{1}^{+}$ and $t_{2}^{+}$, for different F thicknesses in
Fig. \ref{fluxspang}. The flux of transmitted waves is not bounded by unity
now [see Fig. \ref{fluxspang}(a)]. The sharp feature in the flux of
transmitted wave 1 disappears with decreasing $d$ implying that the standing
wave excitation in F is most efficient when wavelength of the incident elastic
wave matches $d$.

The magnetization dynamics or MEW excitation in F can be detected conveniently
via spin pumping~\cite{Tserkovnyak2002} into a thin ($\sim$ few nms)
platinum film~\cite{Weiler2012,Uchida2011} that converts the spin current into
a transverse charge current via the inverse spin Hall effect
(ISHE).~\cite{Saitoh2006} The spin current density injected into a thin Pt
film~\footnote{The Pt film does not
affect the boundary condition considerations when much thinner than the
inverse of the wavenumber that amounts to a few $\mu$m in the GHz regime.} contact on F reads:~\cite{Tserkovnyak2002}
\begin{equation}
\mathbf{J}^{s}=\frac{g_{r}\hbar}{4\pi M_{s}^{2}}\left.  \left(  \mathbf{M}\times
\dot{\mathbf{M}}\right)  \right\vert _{x=d},
\end{equation}
where $g_{r}$ is the real part of the spin mixing conductance per unit
area,~\cite{Tserkovnyak2002} and we disregard its imaginary part as well as
spin current backflow.~\cite{Jiao2013} The time-averaged spin current is
polarized along $\hat{\mathbf{z}}$:
\begin{equation}
\left\langle \mathbf{J}^{s}\right\rangle _{t}=\left\langle \frac{g_{r}\hbar}{4\pi
M_{s}^{2}}\left.  \left(  M_{x}\dot{M}_{y}-M_{y}\dot{M}_{x}\right)
\right\vert _{x=d}\right\rangle _{t}\hat{\mathbf{z}}\ =  \frac{g_{r}\hbar\omega
}{4\pi}\beta^2 \ \hat{\mathbf{z}},
\end{equation}
where we define the `spin pumping angle'~\footnote{$\beta$ is approximately equal to the
precession cone angle.} $\beta = \sqrt{\Im [ \mathfrak{m}_{x}^{\ast
} \mathfrak{m}_{y}/M_{s}^{2}] } $ as a dimensionless measure
of the pumped spin current, with $M_{x,y}(x=d)=\mathfrak{m}_{x,y}e^{-i\omega t}$ and [Eq. (\ref{psif2})]:
\begin{equation}
\mathfrak{m}_{x,y}=\sum_{l}\chi_{l}[1,2]\left(  t_{l}^{+}e^{ik_{l}d}-t_{l}^{-}\right)  .
\end{equation}
The spin current pumped into the Pt film is converted into a transverse
voltage by the ISHE that can be computed by solving the
spin diffusion equation with the appropriate boundary
conditions.~\cite{Mosendz2010}

\begin{figure}[tb]
\begin{center}
\subfloat[]{\includegraphics[height=80mm]{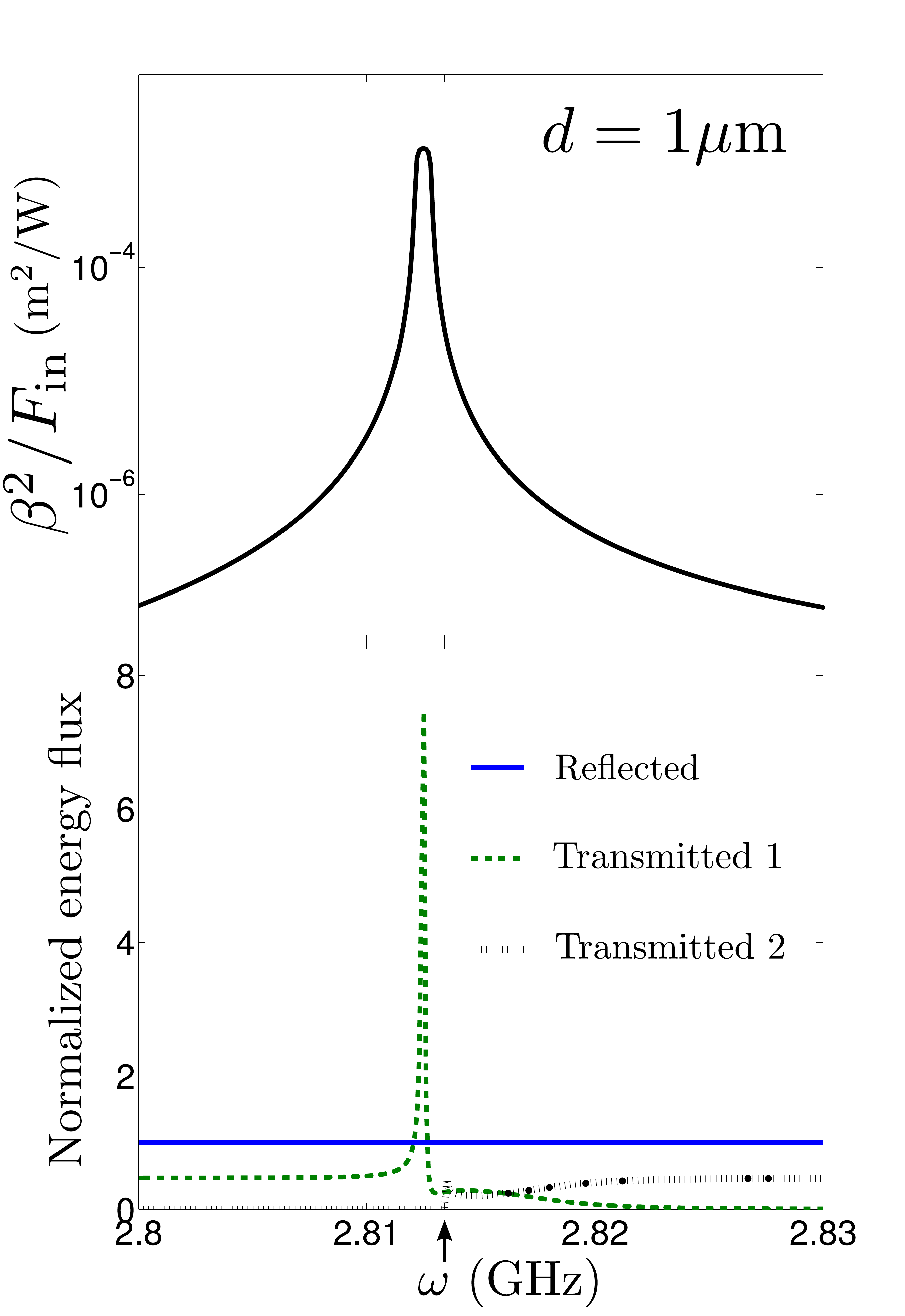}}
\subfloat[]{\includegraphics[height=80mm]{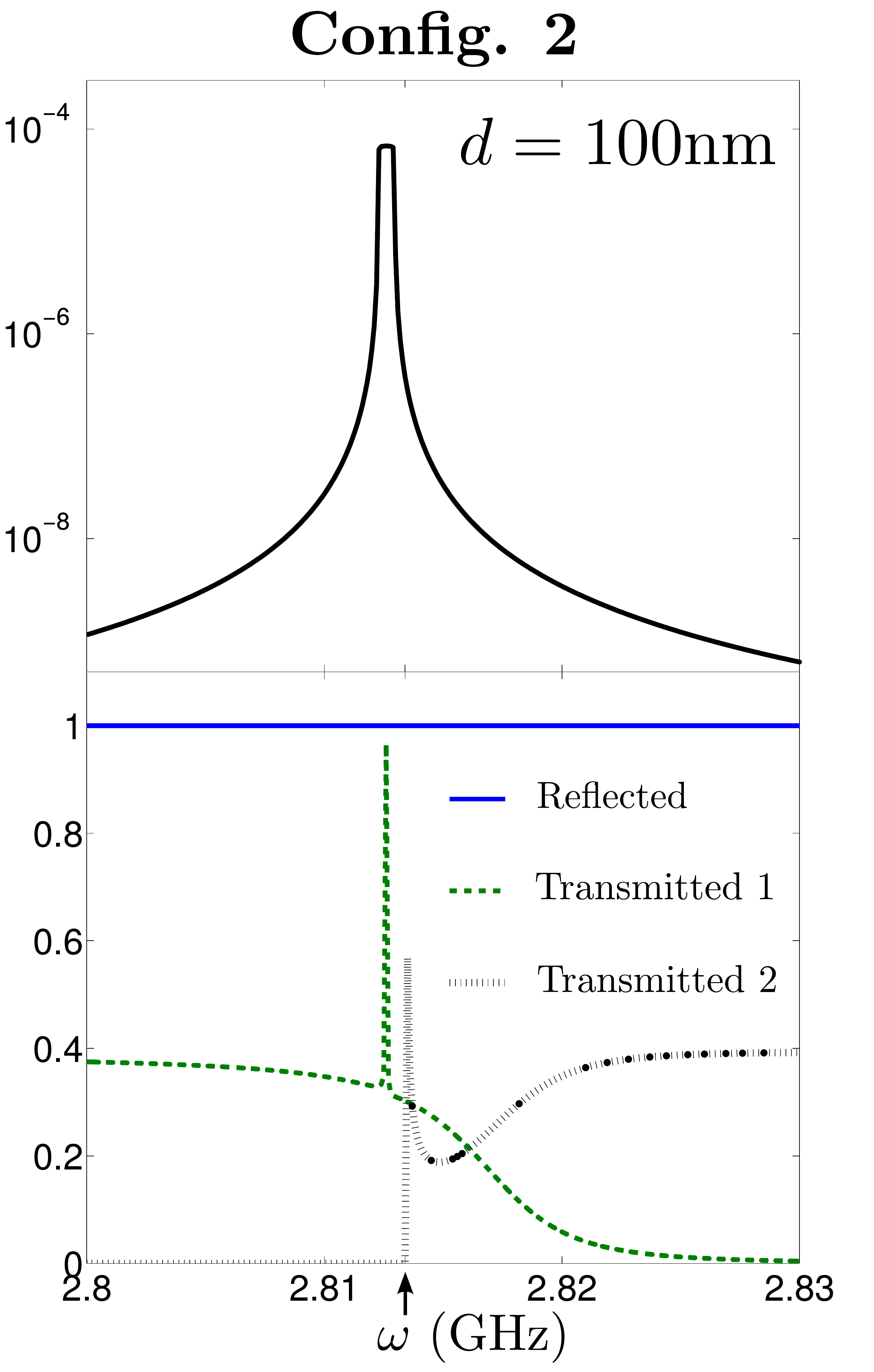}}
\subfloat[]{\includegraphics[height=80mm]{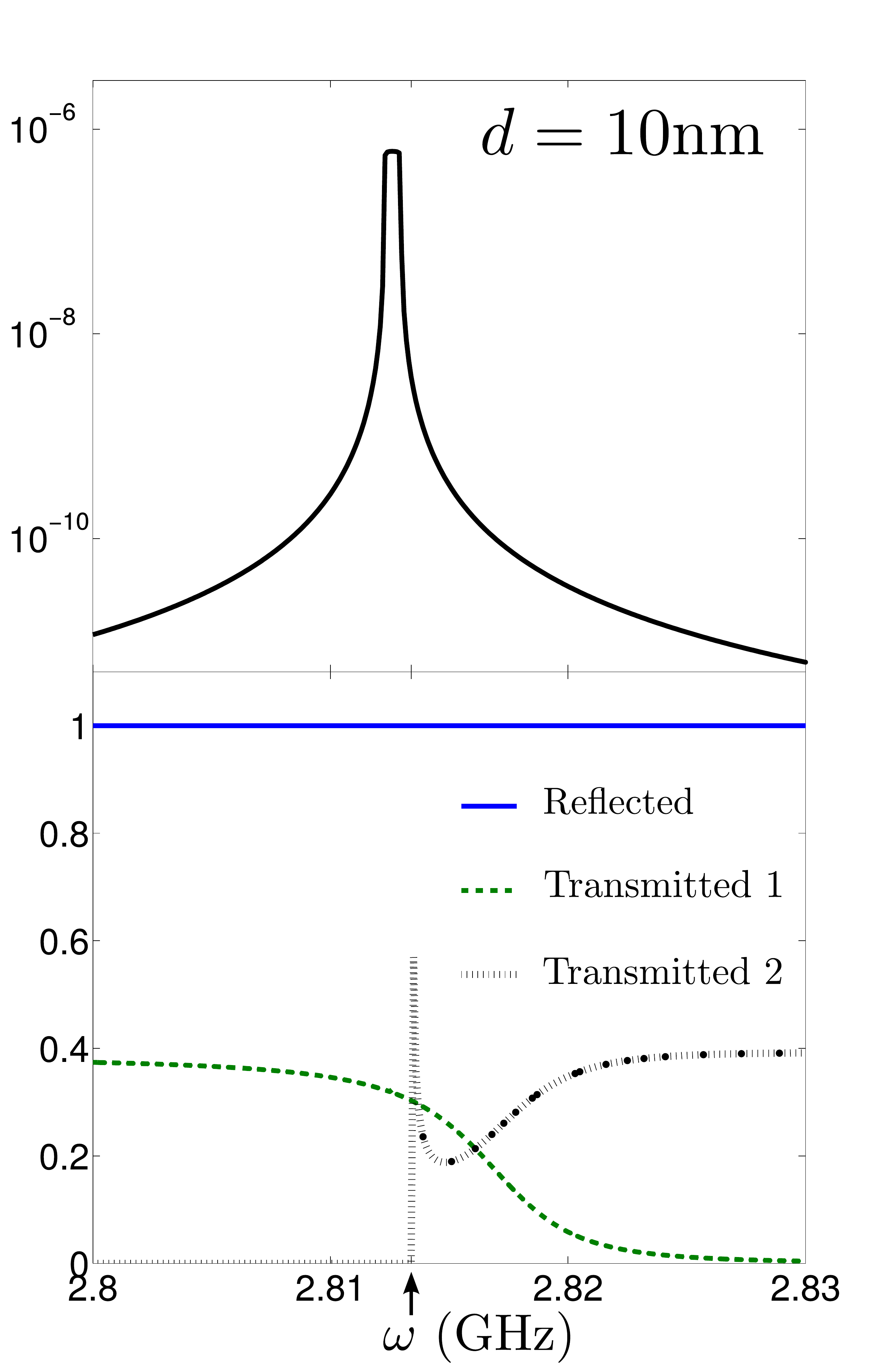}}
\end{center}
\caption[Energy fluxes and squared spin pumping angle per incident flux for different
thicknesses of the ferromagnetic layer in configuration 2]{{\it Configuration 2}: Energy fluxes and squared spin pumping angle
$\beta^2$ per incident flux $F_{\mathrm{in}}$ \textit{vs.} frequency $\omega$ of
the incident elastic wave for F layer thickness (a) $d=1\,\mathrm{\mu m}$,
(b) $d=100\,$nm, and (c) $d=10\,$nm. The arrows on the abscissas indicate the FMR frequency $\omega_{0}$.
The fluxes shown here are carried by the forward propagating transmitted waves
in F and reflected wave in N.}
\label{C2fluxspang}
\end{figure}

The squared spin pumping angle $\beta^2$ is proportional to the incident energy flux
$F_{\mathrm{in}}$. The ratio $\beta^2/F_{\mathrm{in}}$ is plotted against
$\omega$ in Fig. \ref{fluxspang} (upper panels) for different thicknesses $d$. The spin
current is resonantly enhanced around the FMR frequency $\omega_{0}$ with a
maximum that decreases with $d$ as expected from the excitation efficiency
(lower panels in Fig. \ref{fluxspang}). A dip in the frequency dependence of the spin pumping angle develops at a frequency slightly below $\omega_0$ with decreasing $d$ (upper panels in Fig. \ref{fluxspang}). This dip is attributed to an enhanced excitation of the evanescent (counter-rotating) $m^-$ mode, which pumps spin current with opposite polarity, when $d$ is comparable to or less than the decay length (a few hundred nm) of this mode. In Config. 2, the $m^-$ mode does not couple to the incident $r^+$ wave, hence the $\beta^2$ spectra are almost symmetric Lorentzians  (see Fig. \ref{C2fluxspang}).

\begin{figure}[tb]
\begin{center}
\subfloat[]{\includegraphics[height=60mm]{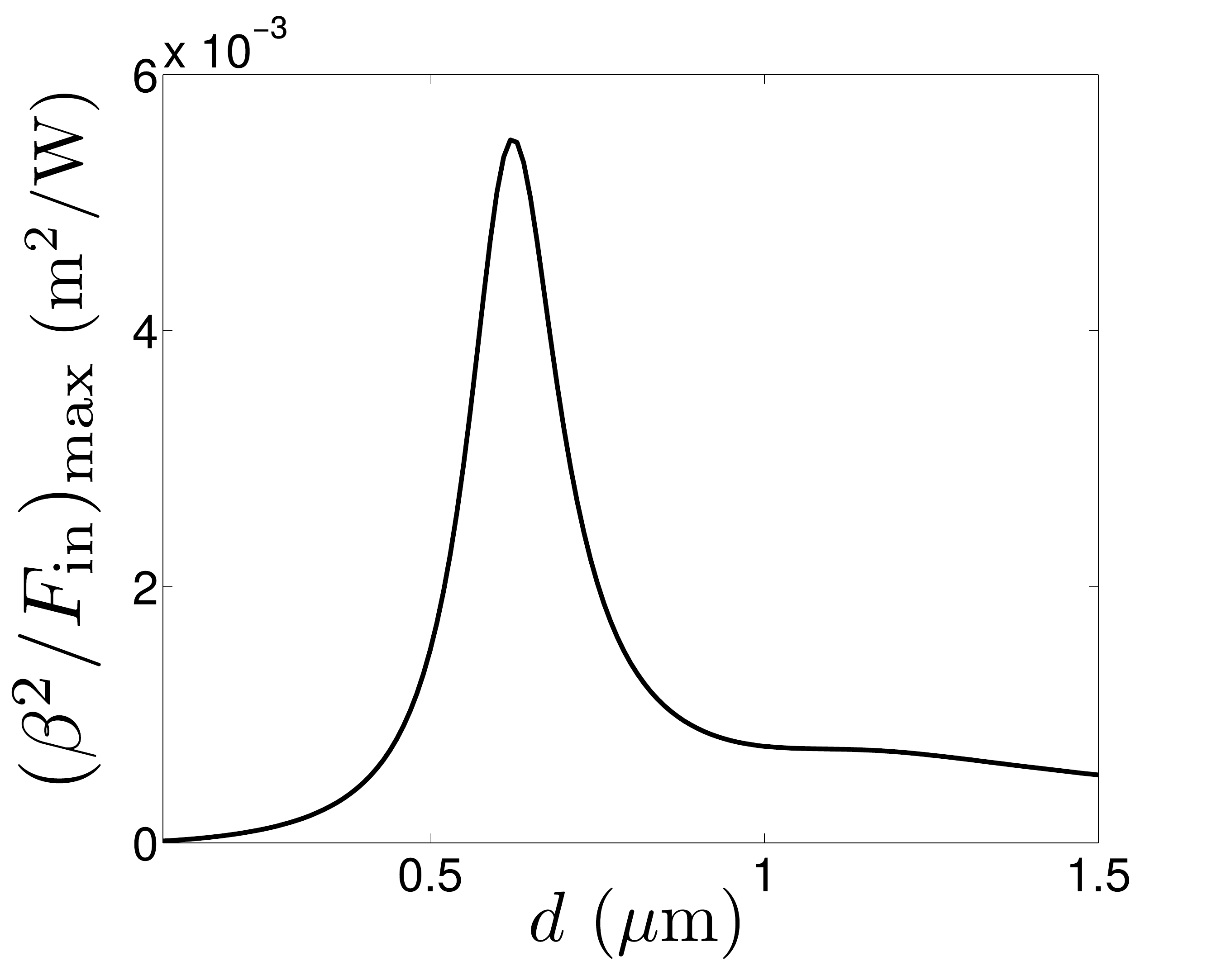}}
\subfloat[]{\includegraphics[height=60mm]{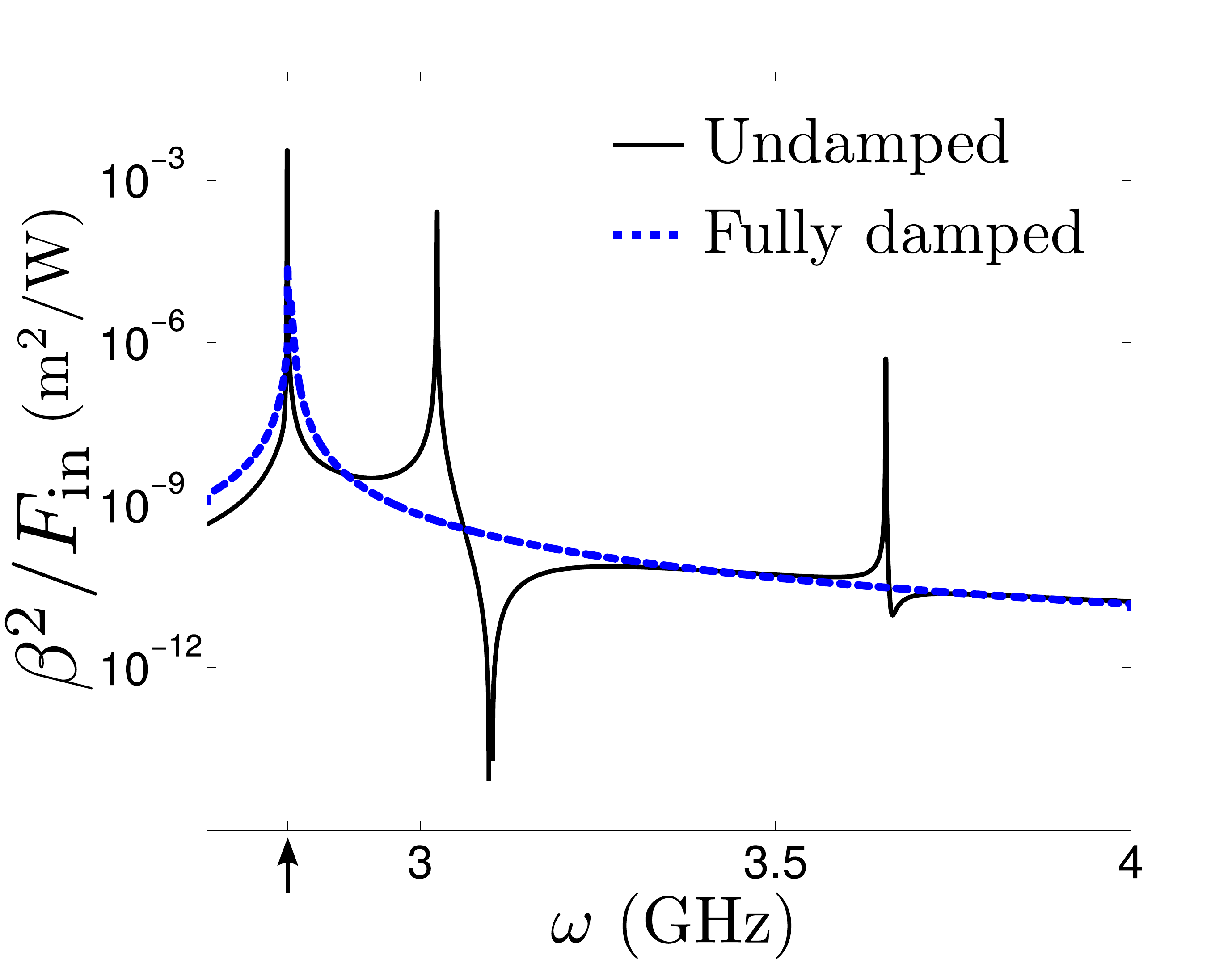}}
\end{center}
\caption[Spin wave excitation by elastic waves]{(a) Maximum value of
$\beta^2/F_{\mathrm{in}}$ in the frequency range around $\omega_{0}$
\textit{vs.} the thickness of the ferromagnetic film ($d$). (b) $\beta^2
/F_{\mathrm{in}}$ \textit{vs.} $\omega$ for $d=0.62~\mu$m corresponding to the
maximum in (a). The peaks corresponding to the first two standing MEWs, in
addition to the uniform mode, can be seen. The fully damped case,
corresponding to an ideal acoustic sink at the far end (or an infinitely thick
F layer as considered in Section \ref{1int}), is depicted by the dashed line.
The arrow on the abscissa indicates the FMR frequency $\omega_{0}$.}%
\label{swexcitation}%
\end{figure}

The maximum value of $\beta^2/F_{\mathrm{in}}$ around $\omega_{0}$ as a function of $d$ in Fig.
\ref{swexcitation}(a) shows a peak at $d\approx0.62\,\mathrm{\mu m}$, a thickness comparable to the wavelength of the incident elastic wave. $\beta^2/F_{\mathrm{in}}$ is plotted for $d=0.62\,\mathrm{\mu m}$ over a wider
frequency range in Fig. \ref{swexcitation} (b). Two additional peaks can be
attributed to spin wave resonances ($k_{n}=n\pi/d,~n=1,2$). A perfect energy
sink at the outer interface, as considered in the previous subsection,
suppresses any reflection. The resulting average squared spin pumping angle per incident flux, depicted
by the blue dashed line in Fig. \ref{swexcitation} (b), is indeed considerably
smaller than in the case of a reflecting interface.

We note that all the excited modes are dominantly magnonic because the
frequencies corresponding to the wavenumbers $k_{n}$ lie in the W region [Fig.
\ref{disp}(b)]. The translational symmetry breaking at the interface allows
excitation of spin waves without wavenumber conservation. $(\beta^2
/F_{\mathrm{in}})_{\mathrm{max}}$ (and hence the spin current) decreases with
increasing $n$.


\section{Conclusion}

\label{conc} We study the excitation of magnetization dynamics in a
ferromagnet (F) via elastic waves injected by an attached non-magnetic
transducer (N). To this end, a scattering theory formulation of the
magneto-elastic waves (MEWs) resulting from magneto-elastic coupling (MEC) in F has been employed. We solve the equations of motion for MEWs propagating orthogonal to (Config. 1) and along (Config. 2) the
equilibrium magnetization direction. Config. 1 leads to excitation of evanescent counter-rotating spin waves, in addition to the two traveling
quasi-spin and quasi-elastic waves. The evanescent waves are not important for energy transport but play significant roles in other
phenomena such as transients in pulsed excitation or evanescent-wave mediated
coupling between two media.~\cite{Uchida2013}

Acoustic excitation of MEWs can efficiently generate magnetization dynamics in the form of magnon-polarons (MPs) around the anti-crossing region.  In sufficiently thin ferromagnetic films standing spin waves can also be excited. The efficiency is
maximized for F layer thicknesses that match the wavelength of the elastic waves. The magnetization dynamics can be detected by spin pumping into an adjacent normal metal layer via the inverse spin Hall effect. The formulation of energy and spin transport by MEWs unifies phononics and magnonics thereby paving the way into yet unchartered territory. 

\section*{Acknowledgments}

A.K. thanks M. Weiler, S. Goennenwein and H. Huebl for fruitful discussions.
This work was supported by the FOM Foundation, Reimei program of the Japan Atomic Energy Agency, EU-FET ``InSpin'', the ICC-IMR, DFG (Germany) Priority Programme 1538
``Spin-Caloric Transport'' (BA2954/2), and JSPS Grants-in-Aid
for Scientific Research (Grant Nos. 25247056, 25220910, 26103006).

\bibliography{meclinear}

\end{document}